\documentstyle[aaspptwo]{article}
\begin{document}
\title{A new chemo-evolutionary population synthesis model for early-type galaxies. I: Theoretical basis}
\author{A. Vazdekis{$^{1}$}}
\author{E. Casuso{$^{1}$}}
\author{R. F. Peletier{$^{2,1}$}}
\author{J. E. Beckman{$^{1}$}}
\affil{$^{1}$Instituto de Astrofisica de Canarias,E-38200 La Laguna,
Tenerife, Spain}
\affil{$^{2}$Kapteyn Instituut, Postbus 800, 9700 AV Groningen, Netherlands}
\journalid{Vol}{Journ. Date}
\articleid{start page}{end page}
\paperid{manuscript id}
\cpright{type}{year}
\ccc{code}
\lefthead{Vazdekis, Casuso, Peletier \& Beckman}
\righthead{A new chemo-evolutionary population synthesis model for early-type galaxies}

\begin{abstract}
We have developed a new stellar population synthesis model
designed to study early-type galaxies. It provides optical
and near-infrared colors, and line indices for 25 absorption lines. 
It can synthesize single age, single metallicity stellar 
populations or follow the galaxy through its evolution from an initial gas 
cloud to the present time. The model incorporates the new isochrones
of the Padova group and the latest stellar spectral libraries.
We have applied our model to new data for a set of three early-type galaxies, 
to find out whether these can
be fitted using single-age old metal-rich stellar
populations, as is normal practice when one uses other stellar models 
of this kind. The model is extensively compared with previous 
ones in the literature to establish its accuracy as well as the 
accuracy of this kind of models in general. 

Using the evolutionary version of the model we find that we cannot fit the most
metal-rich elliptical galaxies if we keep the IMF constant and
do not allow infall of gas. We do however reproduce the results
of Arimoto \& Yoshii (1986) for the evolution of the gas, and
produce colors, and, for the first time with this 
type of models, absorption line-strengths. It is in fact possible to fit the
data for the elliptical galaxies by varying the IMF with time.
Our numerical model is in good broad agreement with the analytical 
{\em simple model}. We prefer however to calculate the evolution
of the gas numerically instead of using the {\em simple model}, since it offers
 more flexibility, and even improved insight, when comparing with observations. In the present paper we describe the model, and compare a few key
observables with new data for three early-type {\em standard} galaxies. 
However the data, as well as our fits, will be discussed in much more detail in
a second paper (Vazdekis {\it et al.} 1996), where some conclusions will be 
drawn about elliptical galaxies on the basis of this model.
\end{abstract}


\section{INTRODUCTION}
\subsection{Population Synthesis}
The information that is available about galaxies primarily relates to their 
morphology, internal kinematics, and spectral energy distribution. The further away one goes, the less important become the first two, compared to the third.
 Clearly, if one wants to study galaxy evolution, study of the electromagnetic 
spectrum is of maximum importance. Since the spectrum of a galaxy generally consists of a combination of stellar spectra, emission from gas, and possibly non-thermal radiation, partly extinguished by dust, and since also stellar 
spectra exist in a very large number of varieties, it is clear that understanding the spectrum of a galaxy is very difficult.

To study complicated spectra it is necessary to first understand the 
spectral energy distributions of relatively simple objects. For this reason
we will discuss here a model that analyzes the spectra of early-type galaxies.
These objects appear to contain relatively little dust extinction and gaseous interstellar medium, and to have little recent star formation. 
They have, for these reasons, 
been the most studied objects in the current literature on population synthesis. There have been a number of accepted ways to attack the problem of understanding the spectrum of an elliptical. To understand why we adopt our current method, we
 will summarize shortly some of the  population synthesis methods most often 
used in the literature.

A stellar population synthesis program tries to find a combination of 
stars for which the integrated spectrum agrees with the observed spectrum
of the object under study. In practise the problem is often underconstrained,
i.e. a number of combinations of stars can be found which are able to fit the spectrum.
To overcome this problem one generally forces the solution to obey certain
constraints. These range from simple continuity requirements (e.g. 
the luminosity function should decrease monotonically) to the requirement
that the distribution of stars is determined completely by stellar evolution
calculations. Models with very few physical constraints are generally called
{\em empirical} population synthesis models, as opposed to
{\em evolutionary} models.

Empirical models have been used with some success by Spinrad \& Taylor (1971), Faber (1972), O'Con\-nell (1976, 1980) and Pickles (1985). 
These papers often make use of linear programming to obtain their results. 
Some workers, notably Bica (1988), have attempted to take into account evolutionary effects by using, as units of population, distributions of stars observed in clusters of our Galaxy, instead of individual stars.

Evolutionary models use a theoretical isochrone or HR diagram, convert isochrone parameters to observed spectra in some way and, finally, integrate along the isochrone. They all need to make an assumption about which {\em initial mass
function} IMF to use. Also, the models need a recipe prescribing when the stars
have been formed. Since the IMF is not very well known at the present time, its 
treatment is not very different from one model to another. However, as far
as the {\em star formation rate} (SFR) is concerned, some models assume that all stars are formed at the same time, others prescribe that the SFR has to
decrease exponentially with time, while still others explicitly try to describe the whole formation of a galaxy from a gas cloud and form stars when the physical conditions in the gas are adequate. 
Examples of this kind of evolutionary models can be found in 
Tinsley (1968,1972,1978a,1978b,1980), Searle {\it et al.} (1973), Tinsley \& 
Gunn (1976), Turnrose (1976), Whitford (1978), Larson \& Tinsley (1978), Wu {\it et al.} (1980), VandenBerg (1983), Bruzual (1983,1992), Stetson \& Harris 
(1988),  Renzini \& Buzzoni (1986), Rocca-Volmerange \& Guiderdoni (1987,1988,1990), Guiderdoni \& Rocca-Volmerange (1987,1990,1991), Yoshii \& 
Takahara (1988), Rocca-Volmerange (1989), Buzzoni (1989),  Charlot \& Bruzual (1991), Lacey {\it et al.} (1993) and Bruzual \& Charlot (1993). Some of these
 models not only predict colors but also line-strengths notably those of Peletier (1989) and Worthey (1994).  

There are also models which combine evolutionary population predictions with considerations of chemical evolution. These models follow the evolution of the gas and make use of isochrones of more than one metallicity (solar). Examples of these {\em 
chemo-evolutionary} population synthesis models are Arimoto \& Yoshii (1986,1987), Casuso (1991), Bressan {\it et al.} (1994) as well as the model presented in this paper. 
Note that for this type of models only the global metallicity, Z, is normally taken into account to determine the stellar populations. 
However it is in principle possible to follow the abundance 
distribution separately for each of several important elements in the gas. Examples of these so called chemical evolution models can be found in Larson (1972), Matteucci 
\& Tornambe (1987), Tosi (1988) and there are many more. However, calculating colors and especially absorption line-strengths in such models has not been attempted up to now. Finally, among this kind of models there are a few which combine chemical evolution with dynamics, e.g. Theis, Burkert \& Hensler (1992).

One cannot say that one type of model is better than another. In general,
more observables and physical parameters can be calculated if one makes
more assumptions. If no assumptions are made about the physics, as in the
empirical models, one may end up with solutions that are unphysical.
If however wrong assumptions are made, one will not learn anything about the 
stellar evolution history either. We show later in this paper that our results
can be reproduced using the {\em simple} analytical model and possibly with infall.
This means that we could have replaced the part of the model that
deals with the evolution of the gas by some analytical calculations.
However, we have preferred to build up the numerical machinery, since 
it offers much more flexibility, and even improved insight. 
It is clear that our understanding of all these aspects is improving with time, which means that the models that are to be applied can legitemately be 
more and more complicated, and in this context we have developed the evolutionary
population synthesis code presented in this paper. For the reasons mentioned above it should never be used as a static 
black box out from which a theoretical fit to the data is to be taken, but as an evolving tool, which might help in disentangling stellar populations in a composite system.

\subsection{Evolutionary synthesis in general, and its problems}

If one wants to calculate the final spectrum of a galaxy that
has evolved from a gas cloud, one has to integrate over time the
spectra of all stars that are still {\em living} at the current time.
The number of stars formed at a certain epoch is determined by the 
star formation rate. Little is known about this, so many models
give it a prescribed form, and let it decrease e.g. exponentially.
In this paper we assume that the SFR is proportional to the gas 
density (Schmidt 1959). The gas density itself and the 
chemical evolution of the gas is calculated taking into account
the original gas, and the metal-enriched gas that is ejected by stars.
The yields for the various elements needed for this calculation
are not especially well known, and better knowledge would significantly
improve the model. Other factors that may affect the SFR, such as inflow or
outflow, are not considered in the context of the present models.

The stars living at the current epoch contain stellar populations with 
a mixture of ages and metallicities. To calculate the final spectrum
we decompose the stars into single stellar populations (SSP) each of a single
age and metallicity, and calculate their spectra.
To do this, one needs in the first place theoretical isochrones. The 
parameters of the isochrones depend, amongst others, on opacities of ions 
and molecules, and are reasonably well known for the early stages 
of stellar evolution. However for later phases such as the AGB and the Post-AGB,
evolutionary calculations are very complicated, and could still benefit 
from significant improvement. The isochrones are much better for solar
and sub-solar metallicity than for metal rich stars, because the former can be, and have been, tested observationally on globular clusters. In general, the
relative composition of the elements heavier than Helium in the 
models is close to solar; only recently have people included
for example oxygen-enhancement (Vandenberg 1992) or $\alpha$-element enhancement (Weiss {\it et al.} 1995).

The following step is to obtain a spectrum for a star with physical
parameters given by the isochrones.
Model atmospheres needed for this (e.g. Kurucz 1992) appear to be reasonably
reliable in the blue. In the red, molecular opacities make them, for
the time being, much less reliable. For that reason several authors 
(Faber {\it et al.} 1985, Gorgas {\it et al.} 1993 and Worthey {\it et al.} 1994) have developed 
a method that depends much more on observations. They use a grid
of observed stars with various theoretical parameters to 
determine fitting functions that can be used to calculate an absorption
line index for any combination of theoretical isochrone parameters.
The problem however is that these fitting functions at the moment are 
available only for a limited number of absorption lines, since one needs 
many stars of various types and metallicities to make them. 

If one wants to carry out population synthesis, one needs to cover as large a wavelength range as possible, and also at as high a spectral resolution. 
This is to cover many colors and line indices, since in principle every color and absorption line is affected in a different way by the parameters
above: the SFR, the IMF, the metallicity and abundance ratios etc.
In practice a model with complete coverage from UV to near-IR is hard to implement, because of lack of fitting functions, color calibrations etc.,
and the observations are hard to obtain. In this paper we are concentrating on early-type galaxies,
for which high spectral resolution is less important, due to their large velocity dispersions. We have produced model output for colors and lines
that are relatively easy to obtain and reproduce, and that allow us
to separate effects due to various relevant physical parameters.
To some extent we are limited by the availability of libraries of
stars, especially for the fitting functions, so this aspect too, can be 
significantly improved in the future.

The layout of the paper is as follows. In Section 2 we will describe
how we obtain colors and absorption line indices for a single 
stellar population, emphasizing differences from previous studies,
and especially any improvements. In Section 3 we introduce our chemical 
evolutionary model. In Section 4 we apply our model to a limited set of data from three 
standard galaxies, for which the data is presented in an accompanying
paper (Vazdekis {\it et al.} 1996, {\bf Paper II}). We perform the fits 
for an SSP (the {\em static} option) and for a full evolutionary case.
In Section 5 the conclusions are presented. 

Finally, in Paper II, apart from presenting the observational basis of the data-set used here, we carry out a comprehensive fit of the whole set of indices and colors on the basis of the scenarios 
suggested in this paper. In particular, we compare the fits obtained assuming a single-age single-metallicity stellar population or using the full 
chemical evolution model. We also discuss the relations between elements.

\section{A single-age, single-metallicity stellar population}
The stellar content of a model galaxy calculated by our full evolutionary model consists of stars of various ages and metallicities. In practice its output spectrum is the integrated output of a large number of spectra of stellar populations each with a single age and metallicity (SSPs). In effect, the SSPs
 can be thought of as the building blocks of our stellar population model, although in an evolutionary model these SSPs are not independent but linked by
 physical parameters governing star formation, which makes the evolutionary method intrinsically more meaningful. Many authors, e.g. Worthey (1994) 
(hereafter W94), restrict themselves to calculating a single SSP, which is less general, but whose simplicity can give insights in the stellar populations of a galaxy.
 We will refer to a single age single metallicity model as the {\it static} case of our stellar population model. In this section we will explain in detail how 
integrated colors and line strengths are calculated for this {\it static} case.

This model works as follows. Assuming an age and a metallicity it calculates the present distribution of the stars according to the required isochrone (see next subsection) and weights the light of the stars on this isochrone with the 
assumed IMF (as defined in Section 2.2). Then it calculates the integrated 
colors and absorption lines, weighting each individual contributing star by its luminosity (as explained in Sections 2.3 and 2.4). We present the results of our
 model for a grid of ages and metallicities and compare them with the literature.

\subsection{The stellar data}
\subsubsection{General}
Our models are based on stellar evolution theory and require a set of isochrones to predict the distribution of the stars in the HR diagram at a given time. In this work we have adopted the large grids of theoretical isochrones 
provided by Bertelli {\it et al.} (1994) (hereafter BBCFN). Their initial compositions have been chosen according to the empirical law of Pagel (1989) $\Delta Y/\Delta Z=2.5$ and are [Z=0.0004,Y=0.23], [Z=0.001,Y=0.23], 
[Z=0.004,Y=0.24], [Z=0.008,Y= 0.25], [Z=0.02,Y=0.28] and [Z=0.05,Y=0.352]. All
 these sets of isochrones were computed using the most recent radiative opacities by Iglesias {\it et al.} (1992) (OPAL) except those of $[Z=0.001,Y=0.23]$ which are based on models with the radiative opacities by 
Huebner {\it et al.} (1977) (LAOL). These isochrones cover a wide range of ages 
and of masses, from approximately $4\times 10^{6}$ to $17\times 10^{9}yrs$ and 
$0.6$ to $\sim 72~M_{\odot}$ respectively. Another main advantage in the
 application of these sets of isochrones is that they cover all the stellar
 evolutionary stages: from the ZAMS, beyond the red giant tip until the white 
dwarf stage after the planetary nebula phase.  
For more details about these isochrones we refer the reader to BBCFN.

The set of isochrones with $Z=0.001$ can be safely used because the effect of
 varying from LAOL to OPAL opacities at low metallicities has been estimated to be small by Alongi {\it et al.} (1993) (see also BBCFN). 

\subsubsection{Use of the isochrones of BBCFN}
In order to be used in our spectrophotometric stellar population synthesis
models (see next sections), the isochrones must contain the following
quantities: the age, the initial and the present mass along the isochrone, the effective temperature and the gravity. 

Because the electronic file with the isochrones provided by the Padova group,
 instead of the initial mass contains the indefinite integral of the IMF by number over the mass (FLUM), as defined in BBCFN, we first obtain the initial mass by inverting the equation
\begin{equation}
FLUM = \frac{m^{1-\alpha}} {1-\alpha}
\end{equation}
\noindent
where $\alpha=2.35$. Now, from the initial stellar mass and the mass at a given point on the isochrone, we obtain the gas mass fraction ejected by a star of 
mass $m$ and metallicity $Z$ until time $t$: $R_{z}(m,t)$ (see Section 3.2.1).

Since the gravities are not in the electronic data base of the isochrones we have calculated them using:
\begin{equation}
g=(4\pi\sigma G) \frac{m_{isoc}}{L}T_{eff}^{4}
\end{equation}
\noindent
where $\sigma$ and $G$ are the Stefan-Boltzmann and the gravitational constants respectively and $m_{isoc}$ is the mass along the isochrone.

\subsubsection{The very low-mass stars}
As the low-mass limit of the isochrones used is $0.6 M_{\odot}$, we decided to complete them with stars of lower masses. The reason for this is that, even if 
these stars do not individually contribute substantially, their integrated light may not be negligible (see Table~5, above all in the near IR for old systems, in those cases where we require a population model with an IMF with high slope. 

To complete the lower-mass MS of the isochrones of composition Z=0.008, Z=0.02 and Z=0.05 we used the results for low-mass stars, for Z=0.004, Z=0.01, Z=0.02
 and Z=0.03, generously provided to us by Pols (private communication). The corresponding tracks for solar composition are described in Pols {\it et al.} (1996). These tracks have been calculated using a renewed version (Han, Podsiadlowski \& Eggleton 1994) of the evolution program of Eggleton (1973),
 incorporating an improved equation of state and making use of radiative opacity tables from Alexander \& Ferguson (1994) for the low temperatures. 
Since the stars with masses below $0.6 M_{\odot}$ are always on the ZAMS
and show almost no evolution, it is straightforward to incorporate them into the lower MS via the appropriate isochrones. However, for the isochrones with metallicity Z=0.008 we first linearly interpolated the low-mass stars between 
Z=0.004 and Z=0.01. Finally, to complete the low-MS of the super metal rich isochrones (Z=0.05) we were forced to  extrapolate linearly from the low-mass stars of Z=0.02 and Z=0.03. As a test we see that in the theoretical HR-diagram there is
 good agreement between stars above (from BBCFN) and below $0.6 M_{\odot}$.

\begin{figure}
\plotone{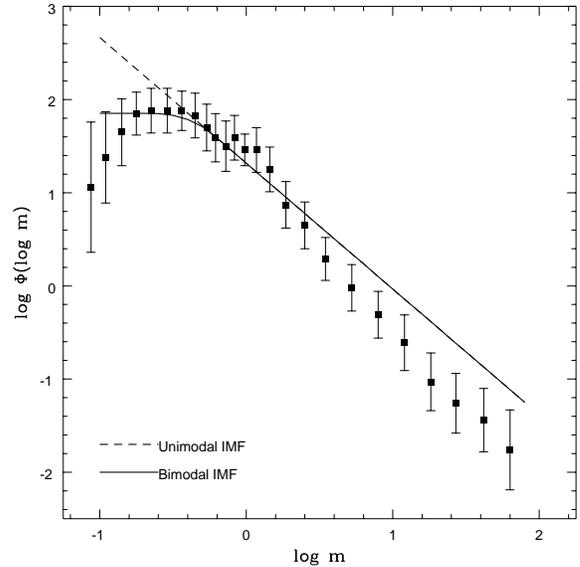}
\caption{Plot of our unimodal and bimodal IMF with slope 1.35. Also represented
are the data of Scalo (1986).}
\end{figure}

\begin{figure}
\plotone{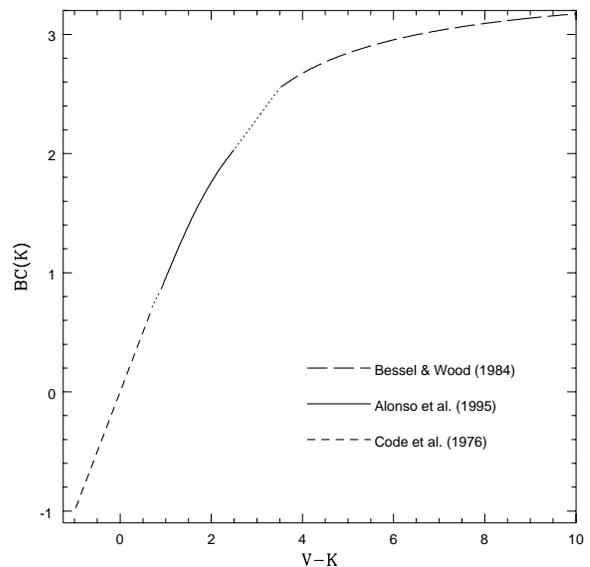}
\caption{Plot of the bolometric correction BC(K) as a function of the color $V-K$ for solar composition. The two dotted portions are linear interpolations to bridge discontinuities at $V-K\sim0.8$ and $V-K \sim3$, which result from using the different sources indicated.}
\end{figure}

\subsection{The initial mass function.}
In this work we adopt two IMF shapes:

\noindent
i) {\bf Unimodal:} 

\begin{equation}
\Phi(m)=\beta m^{-\mu}\\ \hskip1.5cm  
\end{equation}
where $\mu$ for the solar neighborhood is equal to 1.35 ($\mu_{Salpeter}$), and $\beta$ is a constant. 

\noindent
ii) {\bf Bimodal:}
\begin{eqnarray}
\begin{array}{ll}
\Phi(m)=\beta 0.4^{-\mu} & m \leq 0.2 M_{\odot} \\
\Phi(m)=\beta p(m) & 0.2 < m < 0.6 M_{\odot} \\
\Phi(m)=\beta m^{-\mu} & m \geq 0.6 M_{\odot} \\
\end{array}
\end{eqnarray}
where $p(m)$ is a spline for which we calculate the corresponding coefficients solving the following system obtained by the boundary conditions:

\begin{eqnarray}
\begin{array}{l}
p(0.2)=0.4^{-\mu}\\ 
p'(0.2)=0 \\
p(0.6)=0.6^{-\mu} \\
p'(0.6)=-\mu0.6^{-\mu-1}\\
\end{array}
\end{eqnarray}
This bimodal IMF is based on observational results of Scalo (1986) and Kroupa {\it et al.} (1993).

Normalizing our IMF to unity we have 

\begin{equation}
\int_{m_{l}}^{m_{p}}\beta m_{p}^{-\mu}dm+\int_{m_{p}}^{m_{u}}\beta m^{-\mu}dm=1
\end{equation}
\noindent
where $m_{l}$ and $m_{u}$ are, respectively, the lower and upper mass limits of the stellar range considered (in this paper we choose $0.0992 M_{\odot}$ and 72 
$M_{\odot}$ for these limits). We obtain the constant $\beta$ via numerical integration.

In Fig.~1 we plot our IMFs for $\mu=1.35$, comparing them with the data of Scalo (1986).

\subsection{The integrated colors}
The integrated flux of an SSP at a wavelength $\lambda$, $F_{\lambda_{TOT}}$ can be calculated using:

\begin{equation}
F_{\lambda_{TOT}}=\int_{m_{l}}^{m_{T_{G}}}N(m,T_{G})F_{\lambda}(m,T_{G})dm
\end{equation}

\noindent
where $m_{T_{G}}$ is the mass of a star whose lifetime is $T_{G}$, $F_{\lambda}(m,T_{G})$ the flux of a star with mass $m$ and age $T_{G}$, and $N(m,T_{G})$ is the number of these stars. This final distribution of stars is calculated via 

\begin{equation} 
N(m,T_{G})={\Phi (m) \over m}M_{G}\Delta m
\end{equation}

\noindent
where $M_{G}$ is the mass of the galaxy (or of a given region of the galaxy studied). 

For the wavelengths we chose to use the Johnson-Kron-Cousins UBVRI system in the visible and the Bessell-Brett system in the near-infrared. Definitions of the UBV passbands are given in Buser \& Kurucz (1978), for R and I in Bessell (1990) and for JHK in Bessell \& Brett (1988).

To obtain $F_{\lambda}(m,T_{G})$ from the theoretical luminosities, effective temperatures and gravities along the isochrones for each metallicity we preferred to use, whenever possible, empirical relations from observed stars, 
rather than theoretical spectra. We performed this using bolometric corrections (BC) and $T_{eff}$-color and color-color relations. This was 
possible for the isochrones with metallicities not very different from solar: Z=0.008, Z=0.02 and Z=0.05. 
For lower metallicities, due to the lack of calibration stars, we had to take 
the computed colors of BBCFN, obtained by convolving the SEDs of the library of stellar spectra of Kurucz with the theoretical passbands.

\paragraph{Dust analysis.}
To be able to investigate the presence of dust, we have implemented in the code the Galactic reddening law of Rieke \& Lebofsky (1985) which allows us to correct for small amounts of reddening.

\subsubsection{Bolometric corrections}
For early-type stars we calculated the BCs following Code {\it et al.} (1976). For late-type stars we used the formula of Bessell \& Wood (1984) which relates 
the BC to the $V-K$ color. Finally, we used the metal-dependent formulae given in Alonso {\it et al.} (1995) to determine the BCs of the low-MS in the range of temperatures: 4000-8000K. Because the resulting BCs, from different sources, do not match exactly, we have interpolated linearly between them to avoid 
discontinuities. Fig.~2 summarizes the BC for solar composition.

\subsubsection{Temperature-color and color-color conversions}
Where possible we have used empirical studies for which the effective
temperatures had been determined using reliable stellar angular diameter measurements
 (using the lunar occultation technique). For color-color conversions we preferred studies which include a full set of colors rather than those which give a more restricted set, and require the inclusion of subsidiary information to complete 
them. In this way we avoid, as far as possible, problems derived from the use of different sets of filters. Nowadays the various observatories are quite well
 standardized. It is rather a question of different filter sets which were used at different times. Separate conversions were performed for giants and dwarfs:

\begin{itemize}
\item Dwarfs in the range 4000-8000~K: except for U-V (for which we used the color-color relations of Johnson 1966), we inverted the polynomial fitting functions of Alonso {\it et al.} (1996) to obtain each color as a function of 
the $T_{eff}$ and the metallicity. To obtain their relations they observed a large sample of MS stars ($\sim500$) with spectral types from F0 to K5 covering a wide range in metallicity ($-3.0 < [M/H] < 0.5$). To transform from their J 
and K filters to our system we used the relations given in Alonso {\it et al.} (1994). 
\item Stars hotter than 8000~K: we first obtained the $B-V$ from the $T_{eff}$-($B-V$) table of Code {\it et al.} (1976) and then used the color-color relations of 
Johnson (1966). We took a smoothly interpolated transition from Alonso {\it et al.} (1996) for the boundary at 8000~K. This method works since line blanketing for these hot stars is not very important.
For the case of stars hotter than 34000~K we used the colors
given in BBCFN (up to 50000~K obtained by convolving with the Kurucz stellar
 spectra library and above 50000~K by assigning pure black-body (BB) spectra, see BBCFN for details). Finally, to convert from the R and I broad-band filters 
in the Johnson (1966) system to those in the {\it Johnson, Kron, Cousins} system (Bessell 1990) we used the linear relations of Bessell (1983).

\item Dwarfs cooler than 4000~K: due to lack of reliable observations we took the BB effective temperatures of Johnson (1966). The same paper, which does not list 
any metallicity dependence, was used to obtain the different colors by shifting them to match the well-calibrated metal-dependent colors of Alonso {\it et al.} (1996) at 4000~K. In this way we have in fact obtained the dependence on
 metallicity, which becomes more important as the temperature decreases, as can be seen in Alonso {\it et al.} (1996).

\item Giants in the range 3350-4930~K: using the empirical calibration of Ridgway {\it et al.} (1980) we first obtained the ($V-K$) color from $T_{eff}$,
 after which we applied the color-color conversions for giants of Bessell \& Brett (1988) to obtain the other colors. For the cooler giants we also used 
the stellar library of Fluks {\it et al.} (1994). 

\item Giants cooler than 3350~K: unfortunately there is a lack of observations for these stars. Especially for old stellar populations, the contribution from 
giants with spectral types later than M5 is not negligible in red filters especially for steep IMFs. For these stars we used the metal-dependent model
colors of Bessell {\it et al.} (1989, 1991). Since these models do not contain the U and B filters, we obtained the $U-V$ and $B-V$ colors from the observational stellar library of Fluks {\it et al.} (1994), taking stars with the corresponding $V-K$. Finally, to avoid problems in matching these stars 
with hotter giants we have interpolated the colors between 3200 and 3350~K.
We strongly encourage observers to carry out programs to obtain good empirical data for cool giants as well as for cool dwarfs. It is also important to study
 empirically the dependence on the metal content. In this respect observations in Baade's window such as those obtained by Terndrup {\it et al.} (1990 and 1991) are very useful.
\end{itemize}

\paragraph{The low-mass metal-poor stars.}
For stars with masses lower than $0.6M_{\odot}$ and low metallicities (Z= 0.0004, Z=0.001 and Z=0.004), we obtained their parameters by extrapolating linearly
 from the low-MS range of the isochrones for which we took the colors given by the BBCFN. We were forced to this choice due to the lack of empirical data to
 transform from the theoretical to the observational plane. 

\subsection{Line index synthesis}
As well as the colors, we have also calculated absorption line strengths. To make it possible to compare observations with the line strengths from the
models, we have worked in the extended Lick-system (Worthey {\it et al.} 1994). This has the additional advantage that many observed spectra are available for calibration: stars in that paper and also in Faber {\it et al.} (1985) and 
Gorgas {\it et al.} (1993). We used the fitting functions of Worthey {\it et al.} (1994) to relate each line index to its corresponding three atmospheric 
parameters: $[M/H]$, $log\,g$ and $T_{eff}$ (or $V-K$). Since for the CN1 and CN2 indices these authors do not give fitting functions for metallicities below $[M/H] = -1$ we have obtained second order polynomial relations, which 
are given in Table~1. For this we used around $\sim30$ stars of their sample with low metallicities and for which they measured the two indices.  
We also have included some features in the near-IR: the Ca{\sc ii} triplet and the Mg{\sc i} line. We have used the stellar spectra of D{\'\i}az {\it et al.} (1989) to calibrate second order polynomial relations between each line index 
and its corresponding atmospheric parameters. All these fitting functions have been tabulated in Table~1. As pointed out by D{\'\i}az {\it et al.} (1989) it is useful to distinguish between two ranges, metal-poor and metal-rich (separated 
at $[M/H]=-0.3$). In the first the index is strongly dependent on $[M/H]$, while in the second it depends much more strongly on gravity.

We assigned a luminosity class to each {\em live} star by adopting a rather broad set of criteria to distinguish between dwarfs and giants. One of the best discriminators for this is clearly $log\,g$ and we have used as limits:
\begin{eqnarray}
\log\,g \geq 4.0 & {\rm dwarfs} \nonumber \\
3.5 < \log\,g < 4.0 &\left\{\begin{array}{ll}
($V-K$) \leq 2 \log\,g -6 & {\rm dwarfs} \\
($V-K$) > 2 \log\,g -6    & {\rm giants}\\
\end{array}
\right. 
\nonumber\\
\log\,g \leq 3.5 & giants
\end{eqnarray}

\noindent 
Because in the gravity range $3.5 < \log\,g < 4$, the value of $\log\,g$ alone cannot discriminate between the two classes of stars, we decided to use the $V-K$ criterion cited to divide the subgiant region into two parts. Finally, stars 
with $(V-K) < -1.0$ or $\log\,T_{eff} > 4.63$ are not classified, because they are not important for line-strength computations. This equation was mainly used when calculating the contributions of the M stars to the integrated absorption 
features.

The code yields a stellar distribution for a given age of the galaxy and metallicity ($Z$), converted to [M/H] by the relation
\begin{equation}
[M/H]=\log(Z/Z_{\odot})
\end{equation}
\noindent
To calculate an integrated line index the code integrates the contributions of all the stars (obtained using Eq.~11) to calculate the total line index for the resulting SSP. 
An integrated line-index $W_{i}$ is a flux-weighted mean flux and could be expressed either as an equivalent width (EW) or in magnitudes. We have decided 
to maintain the convention of Burstein {\it et al.} (1984), used also in Worthey {\it et al.} (1994) of expressing the molecular-band features in magnitudes and the atomic-line features in equivalent width (EW) in \AA. The NIR features are expressed in EW as well. We calculate $W_{i}$ in the following way: 
\begin{equation}
W_{i}=\frac{\int_{m_{l}}^{m_{T_{G}}}W(m,T_{G})N(m,T_{G})F_{c}(m,T_{G})dm}{\int_{m_{l}}^{m_{T_{G}}}N(m,T_{G})F_{c}(m,T_{G})dm}
\end{equation}
\noindent
where $F_{c}(m,T_{G})$ is the flux in the continuum corresponding to the central wavelength of the spectral feature of the star of mass $m$ at $T_{G}$, obtained by a linear interpolation between the closest broad-band filter fluxes $F_{\lambda_{l}}$ and $F_{\lambda_{u}}$
\begin{equation}
F_{c}(m,T_{G})=F_{\lambda_{l}}+(F_{\lambda_{u}}-F_{\lambda_{l}})\left(\frac{\lambda_{i}-\lambda_{l}}{\lambda_{u}-\lambda_{l}}\right)
\end{equation}
\noindent
where $\lambda_{i}$, $\lambda_{l}$ and $\lambda_{u}$ are the central wavelengths of the index and the two broad-band filters respectively.

Finally, to convert the indices expressed in EW to magnitudes we use the following relation
\begin{equation}
mag=-2.5\log\left(\frac{\Delta \lambda_{i}-W_{i}}{\Delta \lambda_{i}}\right)
\end{equation}
\noindent
where $\Delta \lambda_{i}$ is the wavelength width (in \AA) of the feature.

\subsection{The model input}
To model a single stellar population the main input parameters which must be defined are:
 
\begin{itemize}
\item The IMF shape (unimodal or bimodal as defined in Section 2.2), the IMF slope $\mu$ and the lower and upper cutoff limits $m_{l}$, $m_{u}$ (we kept them fixed at 0.0992 and 72 M$_{\odot}$).
 
\item The age $T_{G}$ of the galaxy (or of the observed zone of the galaxy).

\item The metallicity $Z$.
\end{itemize}

\subsection{The derived colors and absorption lines for a single stellar population}
In Tables~2 and 3 we present the resulting colors and line indices for single stellar populations for unimodal and bimodal IMFs with two slopes $\mu$: 1.35 
and 2.35. For each $\mu$ we present the results for different metallicities and ages. The differences between the two IMFs mainly affect the reddest colors or 
indices. These Tables, as well as others which also include data for models with other $\mu$ values can be obtained from the authors.

\onecolumn
\begin{table}
\hoffset-1cm
\scriptsize
\begin{center}
\begin{tabular}{l|rrrrrrrrrr|l}
\hline
\hline
Index&$g^{2}$&$Z^{2}$&$T_{eff}^{2}$&$g$&$Z$&$T_{eff}$&$Z\,g$&$g\,T_{eff}$&$Z\,T_{eff}$&Constant&Validity \\ \hline\\
CN1&-0.0428&0.4230&-2.6264&-7.6603&16.2199&4.5248&-0.1189&2.1176&-3.9428&19.9478&$3980<T_{eff}<5100$\\
CN1&-0.0089&0.0818&-0.0558&-2.8505&0.0702&-2.9541&-0.0002&0.7670&0.0945&12.1739&$5100<T_{eff}<11100$\\\\
CN2&-0.0196&-0.1867&0.1706&-0.0520&-10.5277&-0.1924&0.0541&0.0734&2.6526&-2.5165&$3980<T_{eff}<5100$\\
CN2&-0.0148&0.0170&0.1875&-2.6252&-0.4784&-4.2434&-0.0485&0.7027&0.2019&13.5325&$5100<T_{eff}<11100$\\\\
\hline \\
Ca$_{II}$1&-0.0065&0.0024 & 0.4254& 5.5058& -6.2485& 1.1452&-0.1162&-1.5381& 1.8636& -7.9084&$Z\leq -0.3$\\
Ca$_{II}$1&0.1269 &-0.8879& 0.6840& 4.9417&  4.2020& 0.5141&-0.3062&-1.6106&-0.7240& -7.7763&$Z>-0.3$\\\\
Ca$_{II}$2&0.1669 &-0.1111& 0.6564& 9.9490& -8.8215& 3.8129&-0.1269&-3.0427& 2.6914&-17.0107&$Z\leq -0.3$\\
Ca$_{II}$2&0.2208 &-2.3469& 0.0913& 1.1574& -9.3570& 1.0685&-0.3788&-0.7946& 3.2014&  1.9801&$Z>-0.3$\\\\
Ca$_{II}$3&0.1335 &-0.1789& 1.6471& 9.8307& -2.5977&-4.5822&-0.0627&-2.9351& 0.8503& -0.6660&$Z\leq -0.3$\\
Ca$_{II}$3&0.1801 &-1.6372& 0.6482&-1.1326&-20.1575&-3.4841&-0.6283&-0.1215& 6.2543& 10.3075&$Z>-0.3$\\\\
Mg$_{I}$  &0.0425 &0.0453 & 0.5193& 3.9973&  1.9597&-3.2429&-0.0278&-1.1445&-0.4173&  5.9036&$Z\leq -0.3$\\
Mg$_{I}$  &0.0302 &0.2442 &-0.8184& 0.9179&  1.3805& 4.0668& 0.0241&-0.2869&-0.3096& -3.1040&$Z>-0.3$\\
\hline
\end{tabular}
\caption{The fitting functions for the CN1 and CN2 indices (as defined in Worthey et al. (1994) for $[M/H] \leq -1$ and for the near-IR features as 
defined in D{\'\i}az et al. (1989) (who define the region of the continuum as well). Given are the polynomial coefficients where Z should be read as [M/H], g as $log\,g$ and T$_{eff}$ as $log\,T_{eff}$}
\end{center}
\end{table}
\twocolumn

\onecolumn
\begin{table} 
\tiny
\begin{center}
\begin{tabular}{l|rrrrr|rrrrr|rrrrr}
\hline\hline
$\mu$&&&1.35&&&&&1.35&&&&&1.35&&\\\hline
Z&&&0.008&&&&&0.02&&&&&0.05&&\\
Age&1&4&8&12&17&1&4&8&12&17&1&4&8&12&17\\\hline\hline
(M/L)$_V$  &  0.75&  2.65&  3.87&  5.37&  7.0&  0.91&  2.96&  5.39&  7.05&  9.14&  1.16&  3.78&  7.20& 10.3& 12.8\\
U-V        &  0.687&  1.087&  1.302&  1.401&  1.446&  0.839&  1.328&  1.474&  1.647&  1.777&  0.955&  1.611&  1.778&  1.913&  2.063\\
B-V      &  0.476&  0.780&  0.879&  0.917&  0.933&  0.603&  0.891&  0.946&  1.007&  1.049&  0.717&  1.006&  1.058&  1.100&  1.146\\
V-R        &  0.354&  0.491&  0.538&  0.557&  0.574&  0.389&  0.540&  0.569&  0.600&  0.623&  0.433&  0.600&  0.631&  0.661&  0.687\\
V-I        &  0.820&  1.045&  1.140&  1.175&  1.210&  0.835&  1.144&  1.197&  1.257&  1.304&  0.867&  1.247&  1.304&  1.361&  1.411\\
V-J        &  1.716&  1.993&  2.136&  2.150&  2.161&  1.906&  2.287&  2.343&  2.416&  2.440&  1.997&  2.625&  2.584&  2.622&  2.640\\
V-K      &  2.546&  2.843&  3.003&  2.996&  2.988&  2.815&  3.214&  3.271&  3.348&  3.356&  2.966&  3.669&  3.580&  3.597&  3.585\\
CN1        & -0.143& -0.047& -0.017& -0.009& -0.009& -0.090&  0.013&  0.026&  0.043&  0.057& -0.027&  0.081&  0.101&  0.117&  0.144\\
CN2        & -0.081& -0.010&  0.016&  0.021&  0.019& -0.037&  0.047&  0.056&  0.074&  0.089&  0.017&  0.113&  0.133&  0.151&  0.182\\
Ca4227     &  0.380&  0.950&  1.183&  1.328&  1.433&  0.601&  1.278&  1.518&  1.721&  1.884&  0.901&  1.770&  2.021&  2.234&  2.431\\
G-band     &  0.698&  3.748&  4.676&  5.037&  5.317&  1.937&  4.817&  5.398&  5.742&  5.941&  3.341&  5.673&  6.057&  6.179&  6.181\\
Fe4383     &  0.610&  2.958&  3.843&  4.215&  4.531&  1.951&  4.770&  5.365&  5.894&  6.300&  3.594&  6.605&  7.236&  7.614&  8.004\\
Ca4455     &  0.519&  1.066&  1.276&  1.368&  1.442&  0.907&  1.515&  1.655&  1.795&  1.898&  1.382&  2.005&  2.134&  2.233&  2.332\\
Fe4531     &  1.594&  2.596&  2.940&  3.096&  3.228&  2.214&  3.244&  3.482&  3.715&  3.898&  2.877&  3.915&  4.158&  4.339&  4.531\\
Fe4668     &  0.674&  2.237&  2.817&  2.950&  2.969&  2.532&  4.640&  4.910&  5.293&  5.547&  5.029&  7.608&  8.008&  8.315&  8.702\\
H$\beta$   &  4.909&  2.390&  1.853&  1.613&  1.402&  4.156&  2.007&  1.625&  1.357&  1.157&  3.267&  1.656&  1.339&  1.137&  0.982\\
Fe5015     &  2.948&  4.105&  4.521&  4.645&  4.652&  3.965&  5.301&  5.440&  5.645&  5.767&  5.117&  6.443&  6.539&  6.654&  6.772\\
Mg$_{1}$   &  0.019&  0.054&  0.071&  0.081&  0.092&  0.034&  0.089&  0.106&  0.123&  0.136&  0.064&  0.136&  0.155&  0.171&  0.189\\
Mg$_{2}$   &  0.084&  0.151&  0.184&  0.202&  0.216&  0.123&  0.218&  0.247&  0.274&  0.294&  0.172&  0.295&  0.327&  0.353&  0.377\\
Mg$b$      &  1.560&  2.474&  2.958&  3.192&  3.333&  2.056&  3.391&  3.790&  4.052&  4.233&  2.642&  4.316&  4.703&  4.972&  5.123\\
Fe5270     &  1.338&  2.166&  2.448&  2.573&  2.673&  1.975&  2.827&  2.995&  3.176&  3.321&  2.613&  3.447&  3.623&  3.760&  3.918\\
Fe5335     &  1.004&  1.794&  2.062&  2.181&  2.293&  1.670&  2.530&  2.705&  2.884&  3.016&  2.483&  3.377&  3.547&  3.689&  3.830\\
Fe5406     &  0.573&  1.121&  1.317&  1.395&  1.471&  1.001&  1.624&  1.746&  1.885&  1.987&  1.508&  2.183&  2.304&  2.408&  2.524\\
Fe5709     &  0.537&  0.726&  0.785&  0.794&  0.798&  0.751&  0.970&  0.984&  1.014&  1.022&  1.018&  1.219&  1.229&  1.226&  1.243\\
Fe5782     &  0.356&  0.589&  0.653&  0.666&  0.672&  0.583&  0.818&  0.836&  0.882&  0.908&  0.807&  1.046&  1.064&  1.083&  1.130\\
NaD        &  1.306&  2.119&  2.438&  2.703&  2.979&  1.941&  2.994&  3.402&  3.747&  4.093&  2.835&  4.194&  4.678&  5.093&  5.496\\
TiO$_{I}$  &  0.028&  0.034&  0.038&  0.039&  0.041&  0.026&  0.040&  0.042&  0.045&  0.048&  0.024&  0.049&  0.053&  0.057&  0.060\\
TiO$_{II}$ &  0.028&  0.047&  0.055&  0.058&  0.062&  0.025&  0.061&  0.067&  0.074&  0.080&  0.027&  0.089&  0.096&  0.105&  0.112\\
Ca$_{II}$1 &  1.212&  1.324&  1.364&  1.383&  1.406&  1.731&  1.781&  1.772&  1.786&  1.793&  1.813&  1.961&  1.901&  1.887&  1.892\\
Ca$_{II}$2 &  3.389&  3.589&  3.672&  3.714&  3.759&  4.454&  4.576&  4.502&  4.502&  4.455&  4.542&  4.815&  4.592&  4.496&  4.439\\
Ca$_{II}$3 &  2.850&  3.037&  3.114&  3.162&  3.213&  3.865&  3.868&  3.745&  3.714&  3.631&  3.955&  4.088&  3.752&  3.582&  3.477\\
Mg$_{II}$  &  0.491&  0.634&  0.691&  0.733&  0.773&  0.641&  0.786&  0.825&  0.853&  0.873&  0.843&  1.018&  1.038&  1.062&  1.076\\
\hline
$\mu$&&&2.35&&&&&2.35&&&&&2.35&&\\\hline
Z&&&0.008&&&&&0.02&&&&&0.05&&\\
Age&1&4&8&12&17&1&4&8&12&17&1&4&8&12&17\\\hline\hline
(M/L)$_V$  &  4.82& 12.0& 15.5& 19.6& 23.6&  6.06& 14.3& 22.3& 27.0& 32.5&  7.59& 18.3& 30.0& 39.7& 46.2\\
U-V        &  0.668&  1.114&  1.346&  1.465&  1.530&  0.807&  1.334&  1.518&  1.701&  1.842&  0.920&  1.599&  1.801&  1.951&  2.103\\
B-V      &  0.470&  0.800&  0.905&  0.953&  0.978&  0.593&  0.900&  0.975&  1.041&  1.089&  0.703&  1.007&  1.076&  1.127&  1.177\\
V-R        &  0.358&  0.521&  0.574&  0.600&  0.624&  0.397&  0.568&  0.615&  0.651&  0.679&  0.440&  0.624&  0.674&  0.714&  0.744\\
V-I        &  0.832&  1.135&  1.244&  1.302&  1.353&  0.872&  1.235&  1.340&  1.416&  1.480&  0.915&  1.337&  1.453&  1.544&  1.607\\
V-J        &  1.693&  2.113&  2.275&  2.330&  2.377&  1.886&  2.382&  2.513&  2.608&  2.668&  2.007&  2.702&  2.772&  2.867&  2.921\\
V-K      &  2.492&  2.958&  3.139&  3.184&  3.222&  2.756&  3.295&  3.430&  3.532&  3.582&  2.927&  3.714&  3.744&  3.825&  3.860\\
CN1        & -0.151& -0.060& -0.034& -0.031& -0.033& -0.101& -0.002&  0.010&  0.023&  0.033& -0.040&  0.068&  0.088&  0.101&  0.124\\
CN2        & -0.089& -0.024& -0.003& -0.002& -0.006& -0.048&  0.031&  0.039&  0.052&  0.063&  0.002&  0.098&  0.119&  0.133&  0.160\\
Ca4227     &  0.457&  1.139&  1.424&  1.624&  1.774&  0.665&  1.415&  1.735&  1.977&  2.182&  0.927&  1.826&  2.148&  2.405&  2.620\\
G-band     &  0.549&  3.681&  4.557&  4.880&  5.115&  1.840&  4.716&  5.256&  5.546&  5.691&  3.303&  5.603&  5.928&  6.004&  5.974\\
Fe4383     &  0.509&  2.971&  3.836&  4.201&  4.494&  1.817&  4.684&  5.293&  5.780&  6.136&  3.461&  6.488&  7.118&  7.450&  7.786\\
Ca4455     &  0.511&  1.117&  1.343&  1.455&  1.543&  0.890&  1.527&  1.698&  1.843&  1.952&  1.351&  1.986&  2.132&  2.235&  2.330\\
Fe4531     &  1.603&  2.709&  3.084&  3.278&  3.435&  2.200&  3.290&  3.583&  3.826&  4.018&  2.830&  3.899&  4.177&  4.362&  4.545\\
Fe4668     &  0.454&  1.993&  2.532&  2.639&  2.648&  2.250&  4.365&  4.651&  5.001&  5.222&  4.689&  7.316&  7.753&  8.030&  8.383\\
H$\beta$   &  4.891&  2.169&  1.601&  1.318&  1.081&  4.133&  1.860&  1.401&  1.114&  0.894&  3.267&  1.579&  1.209&  0.983&  0.824\\
Fe5015     &  2.771&  3.977&  4.375&  4.488&  4.497&  3.793&  5.129&  5.277&  5.462&  5.565&  4.956&  6.274&  6.378&  6.476&  6.580\\
Mg$_{1}$   &  0.024&  0.068&  0.088&  0.101&  0.115&  0.037&  0.098&  0.121&  0.139&  0.154&  0.061&  0.138&  0.162&  0.180&  0.198\\
Mg$_{2}$   &  0.090&  0.167&  0.203&  0.224&  0.240&  0.125&  0.228&  0.264&  0.292&  0.313&  0.168&  0.298&  0.337&  0.364&  0.388\\
Mg$b$      &  1.575&  2.581&  3.042&  3.262&  3.387&  2.056&  3.444&  3.857&  4.100&  4.262&  2.631&  4.342&  4.749&  5.008&  5.147\\
Fe5270     &  1.324&  2.236&  2.527&  2.667&  2.773&  1.928&  2.827&  3.024&  3.198&  3.332&  2.529&  3.401&  3.595&  3.722&  3.861\\
Fe5335     &  1.015&  1.883&  2.162&  2.298&  2.415&  1.635&  2.529&  2.725&  2.894&  3.013&  2.401&  3.313&  3.491&  3.620&  3.739\\
Fe5406     &  0.572&  1.175&  1.379&  1.472&  1.553&  0.962&  1.612&  1.752&  1.884&  1.976&  1.434&  2.127&  2.255&  2.349&  2.448\\
Fe5709     &  0.480&  0.642&  0.683&  0.673&  0.659&  0.684&  0.869&  0.849&  0.858&  0.843&  0.943&  1.117&  1.089&  1.055&  1.051\\
Fe5782     &  0.337&  0.562&  0.623&  0.633&  0.637&  0.541&  0.769&  0.784&  0.820&  0.837&  0.743&  0.981&  0.993&  1.000&  1.035\\
NaD        &  1.690&  2.895&  3.334&  3.743&  4.121&  2.253&  3.633&  4.301&  4.727&  5.153&  2.981&  4.608&  5.320&  5.842&  6.262\\
TiO$_{I}$  &  0.029&  0.041&  0.047&  0.051&  0.055&  0.030&  0.049&  0.056&  0.062&  0.068&  0.029&  0.057&  0.067&  0.075&  0.080\\
TiO$_{II}$ &  0.031&  0.061&  0.074&  0.082&  0.090&  0.034&  0.079&  0.096&  0.108&  0.120&  0.036&  0.104&  0.124&  0.140&  0.152\\
Ca$_{II}$1 &  1.220&  1.377&  1.424&  1.454&  1.483&  1.718&  1.807&  1.823&  1.844&  1.859&  1.767&  1.936&  1.906&  1.906&  1.917\\
Ca$_{II}$2 &  3.451&  3.780&  3.881&  3.957&  4.024&  4.286&  4.451&  4.392&  4.397&  4.366&  4.324&  4.593&  4.383&  4.296&  4.254\\
Ca$_{II}$3 &  2.936&  3.246&  3.341&  3.422&  3.494&  3.606&  3.618&  3.479&  3.448&  3.375&  3.580&  3.674&  3.298&  3.117&  3.022\\
Mg$_{II}$  &  0.574&  0.786&  0.850&  0.909&  0.960&  0.691&  0.876&  0.944&  0.978&  1.007&  0.889&  1.082&  1.134&  1.170&  1.189\\
\hline
\end{tabular}
\caption{The model observables for SSPs with a unimodal IMF with slopes 1.35 and 2.35. The age is in Gyr.}
\end{center}
\end{table}

\begin{table}
\tiny
\begin{center}
\begin{tabular}{l|rrrrr|rrrrr|rrrrr}
\hline\hline
$\mu$&&&1.35&&&&&1.35&&&&&1.35&&\\\hline
Z&&&0.008&&&&&0.02&&&&&0.05&&\\
Age&1&4&8&12&17&1&4&8&12&17&1&4&8&12&17\\\hline\hline
(M/L)$_V$  &  0.48&  1.60&  2.27&  3.10&  3.97&  0.59&  1.80&  3.19&  4.11&  5.24&  0.74&  2.29&  4.23&  5.97&  7.26\\
U-V        &  0.685&  1.080&  1.293&  1.388&  1.429&  0.838&  1.324&  1.466&  1.638&  1.765&  0.954&  1.608&  1.773&  1.906&  2.055\\
B-V      &  0.474&  0.776&  0.874&  0.910&  0.923&  0.602&  0.888&  0.942&  1.001&  1.041&  0.716&  1.004&  1.054&  1.095&  1.141\\
V-R        &  0.352&  0.487&  0.532&  0.549&  0.564&  0.388&  0.536&  0.563&  0.592&  0.613&  0.431&  0.597&  0.626&  0.653&  0.678\\
V-I        &  0.815&  1.032&  1.122&  1.151&  1.178&  0.830&  1.131&  1.175&  1.230&  1.269&  0.862&  1.235&  1.280&  1.329&  1.372\\
V-J        &  1.708&  1.973&  2.110&  2.113&  2.111&  1.899&  2.270&  2.313&  2.379&  2.391&  1.989&  2.610&  2.551&  2.574&  2.580\\
V-K      &  2.538&  2.821&  2.975&  2.956&  2.933&  2.808&  3.198&  3.241&  3.311&  3.306&  2.957&  3.655&  3.548&  3.551&  3.525\\
CN1        & -0.143& -0.046& -0.015& -0.006& -0.005& -0.090&  0.013&  0.028&  0.045&  0.060& -0.027&  0.082&  0.102&  0.118&  0.146\\
CN2        & -0.081& -0.009&  0.018&  0.024&  0.023& -0.037&  0.047&  0.058&  0.076&  0.093&  0.017&  0.114&  0.134&  0.153&  0.185\\
Ca4227     &  0.376&  0.933&  1.155&  1.286&  1.378&  0.598&  1.266&  1.496&  1.690&  1.842&  0.899&  1.764&  2.008&  2.214&  2.406\\
G-band     &  0.697&  3.752&  4.689&  5.061&  5.353&  1.936&  4.823&  5.414&  5.767&  5.979&  3.342&  5.679&  6.070&  6.201&  6.210\\
Fe4383     &  0.608&  2.955&  3.845&  4.221&  4.544&  1.950&  4.773&  5.376&  5.914&  6.332&  3.594&  6.612&  7.252&  7.640&  8.042\\
Ca4455     &  0.517&  1.059&  1.265&  1.353&  1.423&  0.906&  1.511&  1.649&  1.788&  1.889&  1.382&  2.005&  2.134&  2.233&  2.333\\
Fe4531     &  1.590&  2.581&  2.919&  3.067&  3.190&  2.211&  3.237&  3.471&  3.702&  3.883&  2.876&  3.915&  4.158&  4.340&  4.534\\
Fe4668     &  0.674&  2.247&  2.839&  2.982&  3.014&  2.531&  4.645&  4.921&  5.313&  5.578&  5.029&  7.614&  8.021&  8.337&  8.733\\
H$\beta$   &  4.918&  2.414&  1.886&  1.658&  1.458&  4.162&  2.022&  1.651&  1.389&  1.198&  3.270&  1.663&  1.352&  1.155&  1.004\\
Fe5015     &  2.947&  4.109&  4.532&  4.662&  4.675&  3.964&  5.306&  5.451&  5.663&  5.793&  5.117&  6.448&  6.549&  6.670&  6.795\\
Mg$_{1}$   &  0.018&  0.052&  0.069&  0.078&  0.089&  0.034&  0.088&  0.104&  0.121&  0.134&  0.064&  0.136&  0.154&  0.170&  0.188\\
Mg$_{2}$   &  0.084&  0.149&  0.182&  0.199&  0.212&  0.123&  0.217&  0.246&  0.272&  0.291&  0.172&  0.295&  0.326&  0.352&  0.376\\
Mg$b$      &  1.557&  2.469&  2.955&  3.191&  3.336&  2.054&  3.387&  3.787&  4.051&  4.234&  2.640&  4.313&  4.700&  4.970&  5.122\\
Fe5270     &  1.334&  2.156&  2.437&  2.559&  2.656&  1.973&  2.825&  2.993&  3.176&  3.323&  2.613&  3.449&  3.629&  3.770&  3.932\\
Fe5335     &  1.000&  1.784&  2.050&  2.165&  2.275&  1.669&  2.529&  2.705&  2.887&  3.022&  2.483&  3.380&  3.555&  3.702&  3.848\\
Fe5406     &  0.570&  1.114&  1.308&  1.384&  1.457&  1.000&  1.623&  1.747&  1.888&  1.992&  1.508&  2.186&  2.310&  2.418&  2.539\\
Fe5709     &  0.539&  0.735&  0.800&  0.815&  0.826&  0.754&  0.979&  1.001&  1.038&  1.054&  1.021&  1.227&  1.246&  1.252&  1.276\\
Fe5782     &  0.356&  0.590&  0.656&  0.671&  0.679&  0.583&  0.821&  0.843&  0.891&  0.921&  0.807&  1.050&  1.072&  1.095&  1.146\\
NaD        &  1.281&  2.034&  2.316&  2.534&  2.761&  1.921&  2.934&  3.295&  3.610&  3.920&  2.821&  4.157&  4.610&  4.999&  5.385\\
TiO$_{I}$  &  0.027&  0.033&  0.036&  0.037&  0.038&  0.025&  0.038&  0.039&  0.042&  0.044&  0.024&  0.048&  0.051&  0.054&  0.057\\
TiO$_{II}$ &  0.027&  0.044&  0.052&  0.053&  0.056&  0.024&  0.059&  0.062&  0.068&  0.073&  0.026&  0.087&  0.092&  0.098&  0.105\\
Ca$_{II}$1 &  1.208&  1.315&  1.352&  1.368&  1.387&  1.729&  1.776&  1.762&  1.774&  1.778&  1.811&  1.960&  1.897&  1.879&  1.884\\
Ca$_{II}$2 &  3.379&  3.562&  3.636&  3.666&  3.698&  4.456&  4.582&  4.511&  4.514&  4.468&  4.547&  4.831&  4.616&  4.525&  4.472\\
Ca$_{II}$3 &  2.839&  3.009&  3.076&  3.111&  3.149&  3.873&  3.889&  3.777&  3.755&  3.680&  3.969&  4.123&  3.811&  3.660&  3.568\\
Mg$_{II}$  &  0.484&  0.614&  0.666&  0.699&  0.730&  0.636&  0.775&  0.806&  0.830&  0.844&  0.838&  1.010&  1.023&  1.042&  1.051\\
\hline
$\mu$&&&2.35&&&&&2.35&&&&&2.35&&\\\hline
Z&&&0.008&&&&&0.02&&&&&0.05&&\\
Age&1&4&8&12&17&1&4&8&12&17&1&4&8&12&17\\\hline\hline
(M/L)$_V$  &  1.37&  3.37&  4.30&  5.44&  6.54&  1.71&  3.99&  6.19&  7.45&  8.94&  2.13&  4.99&  8.10& 10.64& 12.29\\
U-V        &  0.650&  1.071&  1.293&  1.401&  1.453&  0.793&  1.304&  1.473&  1.650&  1.785&  0.912&  1.581&  1.773&  1.917&  2.067\\
B-V      &  0.457&  0.773&  0.875&  0.917&  0.935&  0.583&  0.882&  0.948&  1.010&  1.053&  0.697&  0.995&  1.058&  1.104&  1.151\\
V-R        &  0.341&  0.491&  0.539&  0.559&  0.577&  0.382&  0.541&  0.577&  0.608&  0.631&  0.428&  0.602&  0.641&  0.673&  0.698\\
V-I        &  0.777&  1.039&  1.136&  1.173&  1.206&  0.812&  1.140&  1.206&  1.266&  1.311&  0.856&  1.245&  1.316&  1.377&  1.423\\
V-J        &  1.598&  1.960&  2.111&  2.130&  2.143&  1.790&  2.244&  2.320&  2.397&  2.425&  1.899&  2.572&  2.562&  2.608&  2.629\\
V-K      &  2.385&  2.793&  2.966&  2.969&  2.965&  2.653&  3.154&  3.234&  3.319&  3.333&  2.816&  3.593&  3.540&  3.569&  3.564\\
CN1        & -0.150& -0.055& -0.025& -0.018& -0.017& -0.100&  0.002&  0.017&  0.033&  0.046& -0.040&  0.070&  0.094&  0.109&  0.135\\
CN2        & -0.088& -0.018&  0.007&  0.013&  0.012& -0.047&  0.035&  0.047&  0.064&  0.079&  0.002&  0.101&  0.125&  0.143&  0.174\\
Ca4227     &  0.420&  1.024&  1.265&  1.418&  1.524&  0.636&  1.336&  1.607&  1.818&  1.988&  0.912&  1.784&  2.077&  2.312&  2.512\\
G-band     &  0.532&  3.708&  4.632&  5.000&  5.283&  1.835&  4.759&  5.348&  5.679&  5.871&  3.306&  5.640&  6.000&  6.107&  6.102\\
Fe4383     &  0.484&  2.954&  3.848&  4.238&  4.562&  1.807&  4.710&  5.359&  5.888&  6.295&  3.462&  6.528&  7.202&  7.579&  7.958\\
Ca4455     &  0.492&  1.070&  1.285&  1.385&  1.462&  0.879&  1.507&  1.670&  1.814&  1.921&  1.348&  1.984&  2.133&  2.239&  2.340\\
Fe4531     &  1.564&  2.619&  2.975&  3.144&  3.281&  2.178&  3.253&  3.533&  3.775&  3.966&  2.825&  3.898&  4.183&  4.377&  4.571\\
Fe4668     &  0.442&  2.018&  2.597&  2.733&  2.767&  2.232&  4.375&  4.681&  5.060&  5.313&  4.681&  7.335&  7.799&  8.103&  8.489\\
H$\beta$   &  4.985&  2.328&  1.784&  1.536&  1.329&  4.196&  1.960&  1.541&  1.272&  1.073&  3.293&  1.620&  1.267&  1.053&  0.900\\
Fe5015     &  2.755&  3.990&  4.418&  4.553&  4.579&  3.785&  5.155&  5.327&  5.537&  5.667&  4.956&  6.298&  6.423&  6.541&  6.664\\
Mg$_{1}$   &  0.020&  0.059&  0.077&  0.088&  0.099&  0.034&  0.092&  0.112&  0.130&  0.143&  0.059&  0.136&  0.159&  0.176&  0.194\\
Mg$_{2}$   &  0.085&  0.157&  0.191&  0.209&  0.223&  0.121&  0.221&  0.254&  0.281&  0.301&  0.166&  0.294&  0.331&  0.359&  0.382\\
Mg$b$      &  1.540&  2.536&  3.013&  3.244&  3.379&  2.023&  3.408&  3.823&  4.075&  4.246&  2.607&  4.316&  4.720&  4.981&  5.123\\
Fe5270     &  1.287&  2.182&  2.475&  2.611&  2.715&  1.912&  2.820&  3.024&  3.211&  3.360&  2.528&  3.416&  3.627&  3.771&  3.927\\
Fe5335     &  0.976&  1.826&  2.105&  2.236&  2.353&  1.621&  2.527&  2.734&  2.918&  3.054&  2.402&  3.336&  3.537&  3.688&  3.827\\
Fe5406     &  0.545&  1.134&  1.340&  1.429&  1.509&  0.952&  1.612&  1.761&  1.904&  2.010&  1.435&  2.146&  2.292&  2.404&  2.518\\
Fe5709     &  0.500&  0.703&  0.767&  0.781&  0.789&  0.708&  0.934&  0.952&  0.984&  0.995&  0.967&  1.177&  1.188&  1.184&  1.202\\
Fe5782     &  0.335&  0.574&  0.643&  0.660&  0.670&  0.546&  0.792&  0.821&  0.870&  0.899&  0.751&  1.009&  1.039&  1.062&  1.110\\
NaD        &  1.430&  2.348&  2.668&  2.935&  3.186&  2.043&  3.228&  3.718&  4.063&  4.399&  2.849&  4.363&  4.955&  5.398&  5.780\\
TiO$_{I}$  &  0.025&  0.033&  0.036&  0.037&  0.039&  0.025&  0.039&  0.042&  0.045&  0.048&  0.025&  0.049&  0.054&  0.059&  0.063\\
TiO$_{II}$ &  0.023&  0.044&  0.052&  0.055&  0.058&  0.024&  0.060&  0.067&  0.075&  0.081&  0.028&  0.088&  0.099&  0.108&  0.116\\
Ca$_{II}$1 &  1.174&  1.312&  1.355&  1.375&  1.397&  1.685&  1.763&  1.760&  1.776&  1.783&  1.742&  1.918&  1.869&  1.859&  1.868\\
Ca$_{II}$2 &  3.329&  3.590&  3.675&  3.716&  3.755&  4.289&  4.480&  4.423&  4.435&  4.401&  4.361&  4.685&  4.489&  4.408&  4.369\\
Ca$_{II}$3 &  2.810&  3.049&  3.125&  3.171&  3.214&  3.677&  3.744&  3.636&  3.622&  3.558&  3.720&  3.906&  3.596&  3.450&  3.373\\
Mg$_{II}$  &  0.492&  0.663&  0.717&  0.756&  0.791&  0.629&  0.796&  0.839&  0.865&  0.883&  0.837&  1.027&  1.055&  1.080&  1.091\\
\hline
\end{tabular}
\caption{The model observables for SSPs with a bimodal IMF with slopes 1.35 and 2.35}
\end{center}
\end{table}
\twocolumn

\onecolumn
\begin{figure}
\plotone{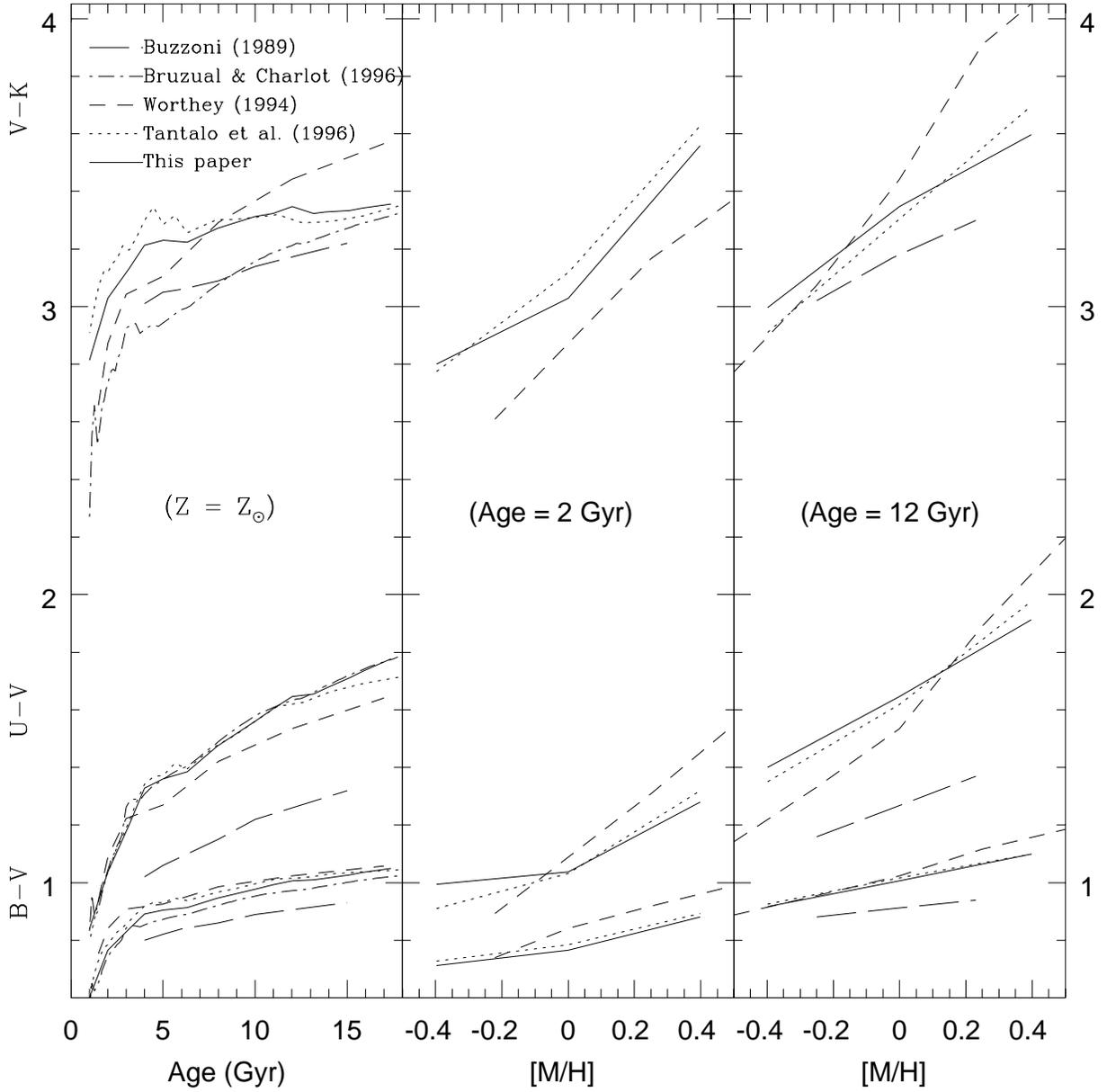}
\caption{Comparison of the colors predicted in our SSP models with those obtained by different authors versus age and metallicity for a Salpeter IMF.}
\end{figure}
\twocolumn

\begin{figure}
\plotone{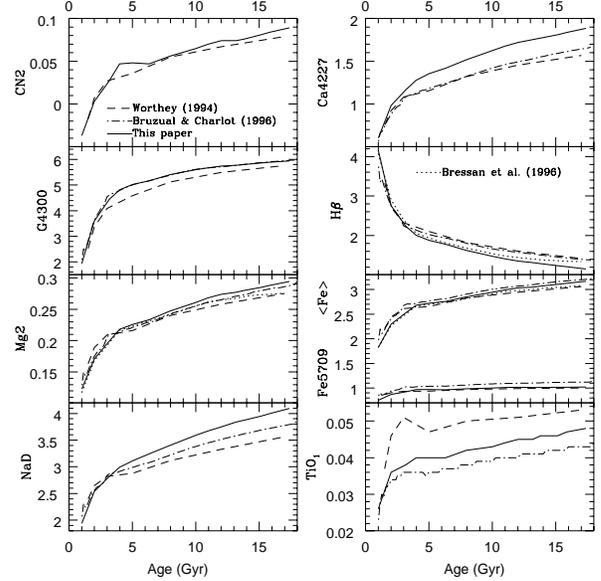}
\caption{A representative set of our synthetic indices versus age for a Salpeter IMF and solar composition compared with the same indices calculated by different authors (from Bressan {\it et al.} (1996) only H$\beta$, Mg$_{2}$ and $<Fe>={Fe5270+Fe5335 \over 2}$ are available).}
\end{figure}

\begin{figure}
\plotone{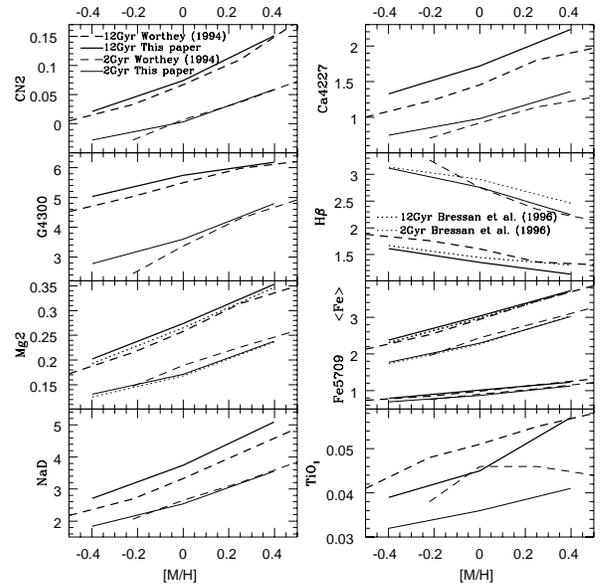}
\caption{The synthetic indices versus metallicity for a Salpeter IMF and for assumed ages of 2 and 12~Gyr, compared with the equivalent plots from W94 and Bressan {\it et al.} (1996).}
\end{figure}

\subsubsection{Comparison with other authors}
As a check on our method we compared our synthetic colors with those published by Buzzoni (1989) (we selected those models which include the red horizontal branch and have a mass-loss rate parameter $\eta=0.3$), Bruzual \& Charlot 
(1996) (private communication, a revised version of the results of Bruzual \& Charlot 1993), Tantalo {\it et al.} (1996) (which includes the revised results of Bressan {\it et al.} 1994) and W94. In Fig.~3 we present the color-age (for solar metallicity) and color-metallicity diagrams for young (2~Gyr) and old 
(12~Gyr) stellar populations. In most of the plots we see that our results are always in better agreement with the values obtained by Tantalo {\it et al.} (1996) and Bruzual \& Charlot (1996), while we differ substantially from the 
results obtained by W94 and Buzzoni (1989). We also note that the older models of the Padova group (Bressan {\it et al.} 1994) give much redder $V-K$ colors than we find (similar to colors of W94) while the model of Bruzual \& Charlot (1993) is much bluer than ours for moderate ages. The similarity between our results and those of Tantalo {\it et al.} (1996) is due to the use of the same 
theoretical isochrones, whereas the small differences are caused by the different conversions to the observational plane. The very red $V-K$ colors of W94 for cool stars are attributed to his bolometric corrections. A nice paper
 analyzing the differences between these models has been written by Charlot {\it et al.} (1996). They found that the scatter between authors is due to the use of
 different isochrones rather than spectral calibrations. Finally, the colors of
 Buzzoni (1989) differ very much from the others. 

For the absorption lines we have compared our results with those of W94, Bruzual \& Charlot (1996) and Bressan {\it et al.} (1996). In Figs. 4 and 5 we investigate the behavior of representative indices, from eight different 
elements. In Fig.~4 these indices are plotted as a function of age for solar composition. In Fig.~5 we plot the same synthetic indices versus the metallicity for young (2~Gyr) and old (12~Gyr) stellar populations. Looking at these figures
we infer that the agreement is generally good. Once again our models agree
better with Bruzual \& Charlot (1996) than with W94. In particular we can notice that our line-strength predictions are in general stronger than the ones of W94,
except for H$\beta$, as expected, and for TiO$_I$. Since most of these
indices are calculated directly from the theoretical isochrone parameters
without any intermediate conversion to the observational plane, as explained 
in Section 2.4, we attribute this to the fact that our isochrones are slightly
cooler, especially for the MS and the turnoff stars, as well as the upper-RGB
and AGB stars. This difference in the temperature is mainly caused by the adopted opacities. However other different stellar evolution 
prescriptions are also significant (see for more details Charlot {\it et al.}
 1996). The disagreement among the models is greatest for NaD, Ca4227 and TiO$_{I}$. This is well understood if we look at the index-($V-K$) diagrams of Gorgas {\it et al.} (1993) and Worthey {\it et al.} (1994) index-($V-K$) plots
for observed stars. The dependence of Ca4227 and NaD on temperature is very
steep (compared with e.g iron indices) for EWs of $\sim1$ and $\sim2$ respectively (where differences in the integrated predictions also become appreciable), especially for dwarf stars. The case of TiO$_I$ is slightly more 
difficult to understand, since for very cool stars this index changes rapidly as a function of temperature for both, dwarfs and giants. Worthey's TiO$_I$ is stronger than ours because his bolometric corrections, the same reason that his $V-K$ is redder. This is clear when one realises that both are strongly dependent on very red stars, for which W94 has a problem with the BCs 
(see Charlot {\it et al.} 1996).

\begin{figure}
\plotone{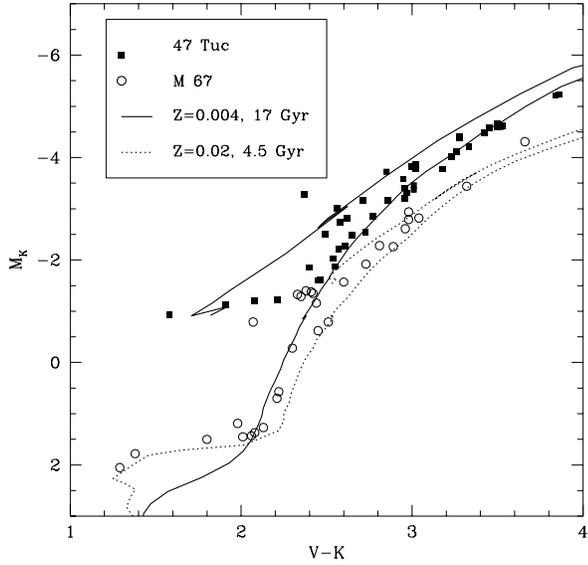}
\caption{Color-magnitude diagram of M~67 and 47~Tuc together with our stellar colors. We plot stars of M~67 from Houdashelt {\it et al.} (1992) and stars of 47~Tuc by Frogel {\it et al.} (1981).}
\end{figure}

\subsubsection{Comparison with observed clusters}
Since clusters are considered to be coeval, they are ideal for testing the 
synthetic colors and absorption lines of our models. In Fig.~6 we have 
compared our model colors and absolute magnitudes with the $M_{K}$ vs. $V-K$ color-magnitude diagrams of Houdashelt {\it et al.} (1992) for M~67
and Frogel {\it et al.} (1981) for 47~Tuc. For M~67 we used an isochrone of 4.5~Gyr and solar metallicity, while for 47~Tuc we used an age of 17~Gyr and Z=0.004. Notice that for 47~Tuc our colors are the same as those of BBCFN, 
since there is no difference in the calibration for low metallicities. 
The fit for M~67 is also excellent, in general the agreement in $V-K$ is better
than 0.05~mag everywhere. W94 obtained a good fit for Z=0.0134 and 4.5~Gyr,
while we use Z=0.02. This difference reflects the difference in temperature of
the isochrones in general.

Next, we made a comparison between our models and the integrated colors as well
as line-strengths of some globular clusters. Two color-color diagrams and $V-K$
vs. Mg$_{2}$ are given in Fig.~7, while in Fig.~8 a number of index-Mg$_{2}$ diagrams are represented. For the colors, the observational data of Galactic and M31 globular clusters are from Burstein {\it et al.} (1984), who used raw data 
from different authors (references therein). We selected, following the
classification of Searle {\it et al.} (1980), UBV data for the LMC globular clusters from Bica {\it et al.} (1992) and Girardi {\it et al.} (1995). The 
line-strengths of Galactic and M~31 globular clusters are from Burstein {\it et al.} (1984) and for other Galactic globular clusters from Covino {\it et al.} 
(1995). The integrated model observables have been obtained with a bimodal IMF of slope $\mu=1.35$ for SSPs with different metallicities and ages (ranging from 1 to 17~Gyr).
In Fig.~8 we see that our models fit the data very well and that the required metallicity is always lower than solar. As one can see in Fig.~8 our models also fit very well the plots of G band, H$\beta$ and iron indices vs. Mg$_{2}$. For 
instance, the H$\beta$-Mg$_{2}$ diagram indicates that the ages of these clusters must be older than say $\sim 10~Gyr$. On the other hand, the NaD and CN1 features do not fit so well. The NaD mismatch can be probably attributed to 
interstellar absorption, which can increase the EW of the observed features by up to 1~\AA~(Gorgas {\it et al.} 1993). However, for the most metal-rich 
clusters, we sometimes observe differences even higher than 1~\AA. Bica {\it et al.} (1991) found similar results for a large sample of galaxies. For the CN lines the mismatch cannot be attributed to extinction since the line is purely 
stellar. Here only for the most metal-rich galaxies is the mismatch large. 
There are all clusters of M31. It could be that abundance ratio variation
effects play a role here in the metal-rich galaxy M~31 (W94). W94 also attributed the CN mismatch observed in the metal-rich galaxy M~31 to abundance
ratios effects. This effect could also be responsible for part of the NaD
mismatch.
\begin{figure}
\plotone{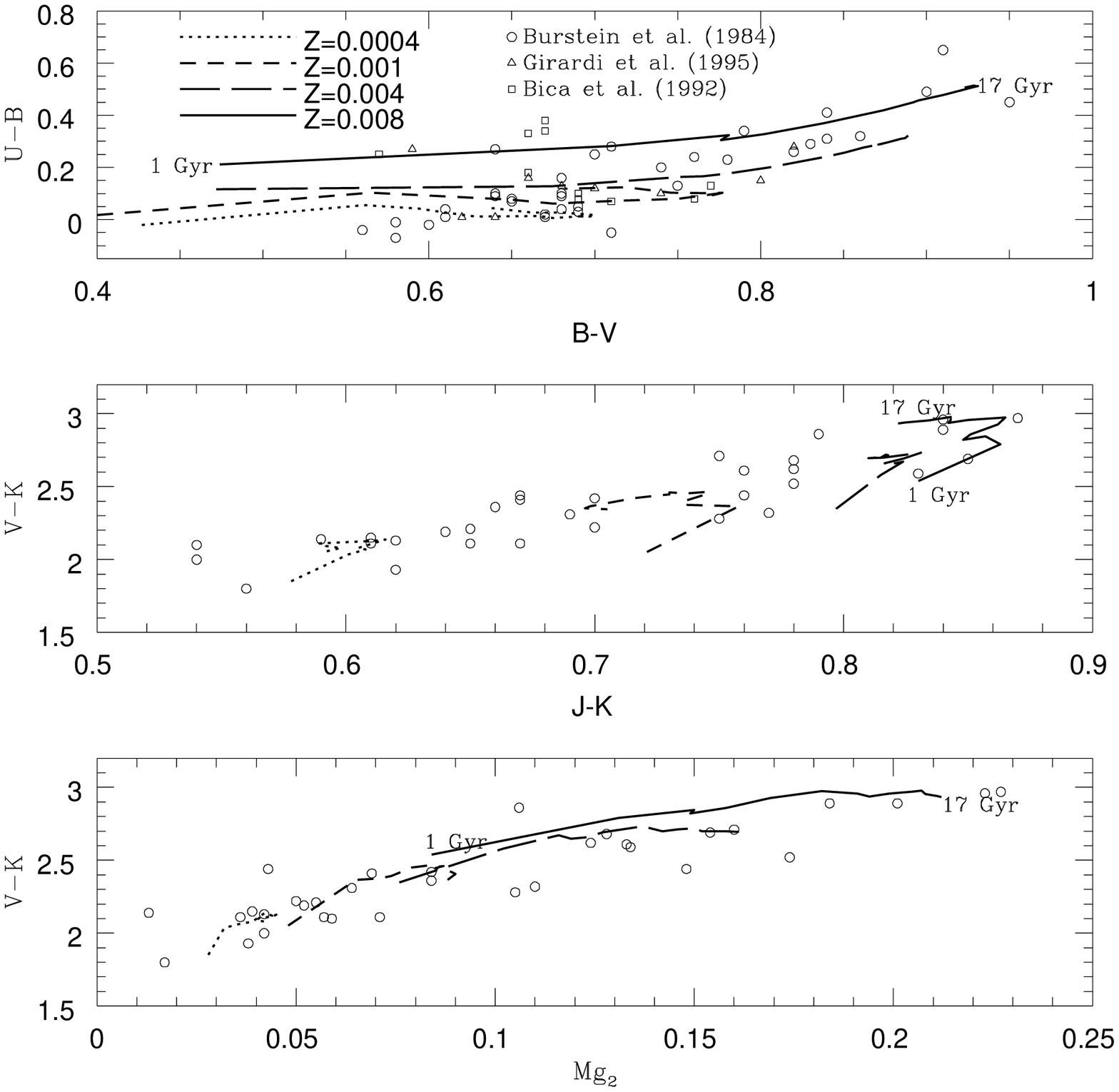}
\caption{Comparison of our models with optical and IR colors and Mg$_{2}$ for Galactic, M31 and LMC globular clusters (see the text for more details). Presented are models with a bimodal IMF of slope $\mu=1.35$ for SSPs of different metallicities, with age ranging from 1 to 17~Gyr.}
\end{figure}
\begin{figure}
\plotone{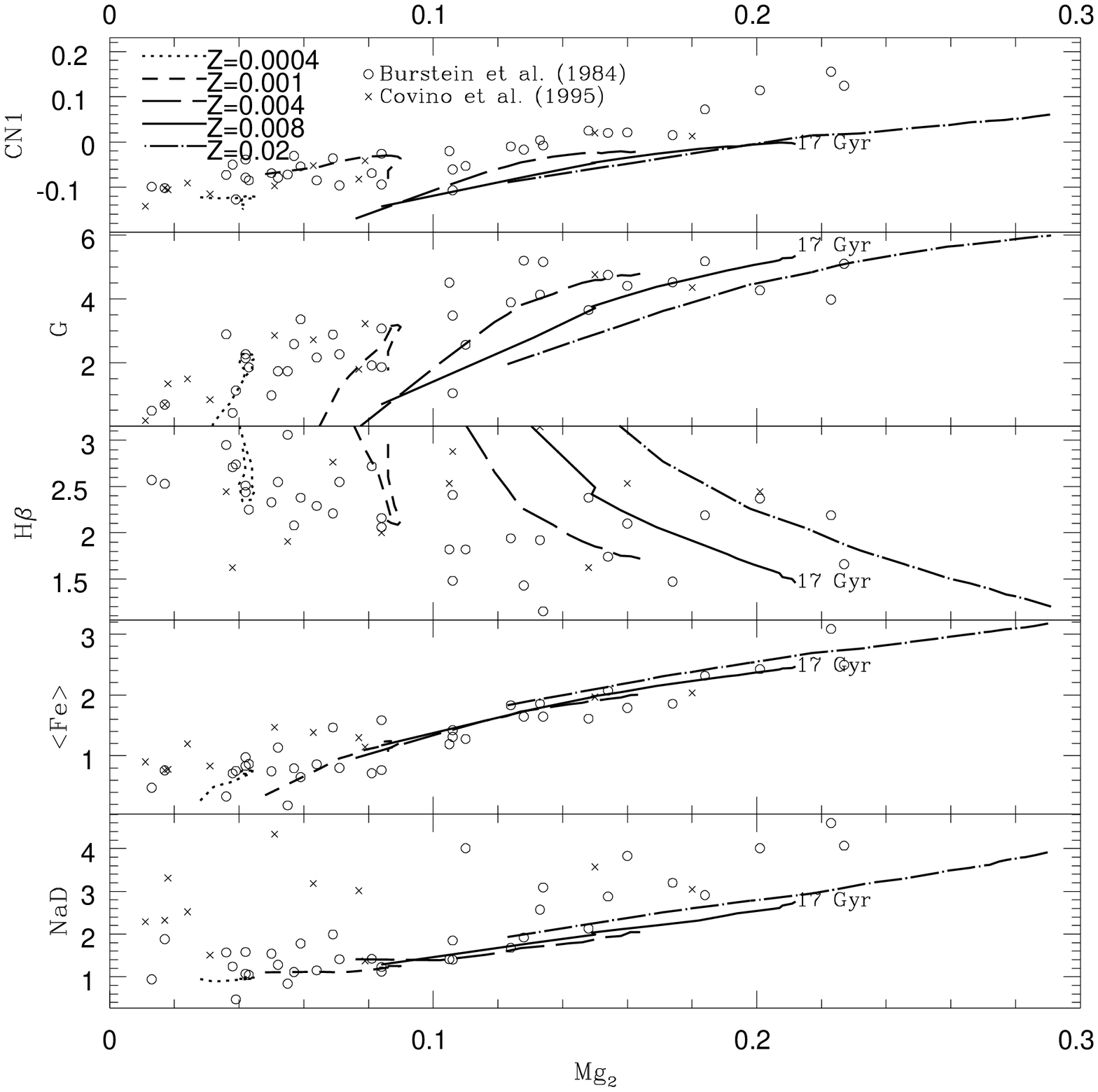}
\caption{Comparison of our integrated indices with observations of Galactic and M31 globular clusters in a number of index-Mg$_2$ diagrams. Our indices are obtained with a bimodal IMF of slope $\mu=1.35$ for SSPs of different metallicities, ranging in age from 1 to 17~Gyr.}
\end{figure}
\clearpage

\begin{figure}
\plotone{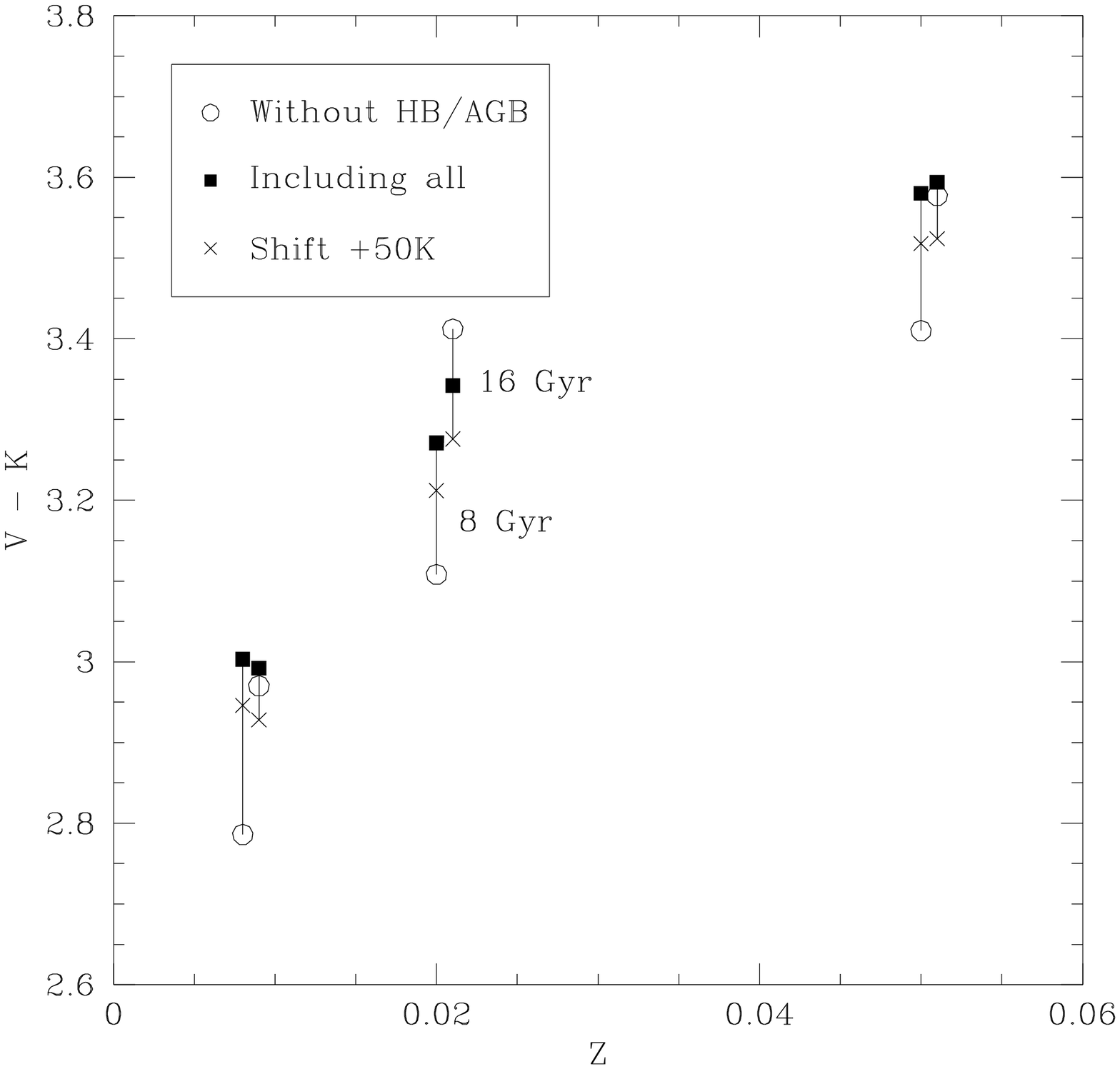}
\caption{$V-K$ for models with and without HB+AGB, and shifting the isochrone by 50~K.}
\end{figure}
\begin{figure}
\plotone{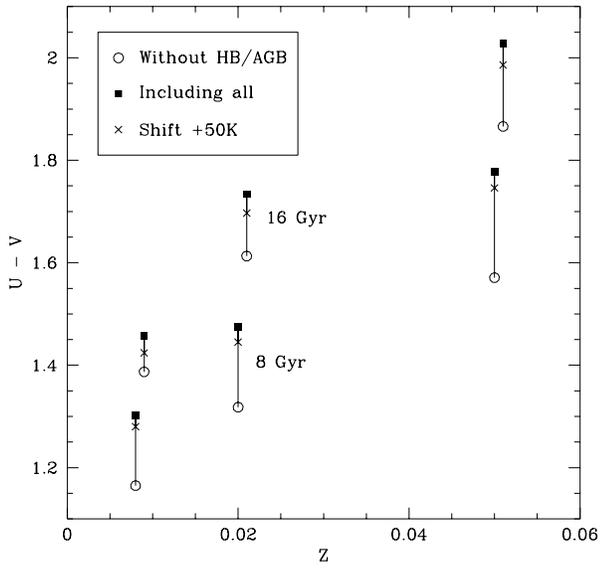}
\caption{$U-V$ for models with and without HB+AGB, and shifting the isochrone by 50~K.}
\end{figure}
\clearpage

\subsection{Errors in integrated line indices and colors}
Since models of this kind are extremely complicated, the determination
of the errors in colors and lines is very difficult. Part of
the physics involved is not completely understood (e.g. molecular opacities),
and even where it is understood calculations can be very difficult 
(e.g. accounting for convection). Errors can be due to uncertainties 
in theoretical stellar parameters, especially for advanced stages of stellar
evolution, to the slope of the IMF, about which little theoretical is known,
to the conversion from the theoretical to the observational plane
using observed and synthetic stars, and to many more factors.

One way to estimate our errors is to compare our results with results from other authors (see Section 2.6.1). But since some errors are systematic, and possibly different for each group, this is not enough. For this reason we also present 
the differences in integrated colors and indices between the values for whole model, and those of all stages {\em except} the HB and the AGB for which the isochrone calculations might contain large uncertainties. In Fig.~9 we present
$V-K$ with and without HB+AGB for various ages and metallicities. Without the
latest phases $V-K$ is quite well-behaved, i.e. monotonic in age and
metallicity. In Fig.~10, we see that the advanced stages contribute 
considerably in $U-V$, which could maybe explain why our models are
amongst the reddest in this color.

The differences are quantified in Table 4. In the last column
we give the maximum differences, in magnitude, or as a fraction of
the line strength. These numbers have the same order of magnitude as
the differences between the authors in Fig.~3.

This table gives an idea of the relative accuracy of the various colors
and lines. The possibilities for systematic errors are considerable. The worst
 colors are $U-V$ and $V-K$, while the models for lines CN$_1$, CN$_2$,
Ca4227, Fe4668 and Fe5782 are the most uncertain. These possible errors have to
be taken into account when applying the model to observational data. Note that
these are not (and do not have to be) the lines for which the comparison between
various authors is the worst.

As an additional test we have shifted the whole isochrone to the blue by 50~K, which might be a reasonable uncertainty in the isochrones. The results are also
plotted in Figs.~9 and 10. As expected, the effect is systematic but does not
change either U-V or $V-K$ by more than $0.05~mag$.

The comparison with M~67 or 47~Tuc shows that here our error in $V-K$ is probably smaller than $0.05~mag$ everywhere. Assuming that the whole RGB (and further 
stages) are $0.1~mag$ too blue (or too red) we find that the integrated $V-K$ is $0.07~mag$ too blue (or too red) for several ages and metallicities. This means that the error due to RGB mismatch is not larger than $0.04~mag$, although at 
Z=0.05 the errors might be larger, since no globular clusters are present here for calibration.

In order to quantify how any mismatch in the calibration of the $K$ vs. $V-K$ diagrams of red giant branches of globulars (see Fig.~6) is propagated through the models, we assumed that the stars on the whole RGB (and further 
stages) are 0.1~mag too blue or 0.1~mag too red.
We then found that the integrated $V-K$ colors are 0.07~mag too blue with respect to the red propagation for models of 17~Gyr, and Z=0.02 and 0.05. For models of 1~Gyr and the same metallicities we obtained a difference of 0.08~mag. This
means that we may assume that the error in integrated $V-K$ up to
the tip of the RGB is smaller than 0.04 mag, although at Z=0.05 this error
could be larger, since no globular clusters here are present to calibrate
the models. At low metallicities there are no differences between our 
integrated colors and those of BBCFN.

\subsubsection{The effect of the very low mass stars on the integrated observables}
In this section we briefly quantify the effects of very low-mass stars ($<0.6M_{\odot}$) on the integrated observables. Since their effects are more
important in the red we have calculated the effects on $V-K$ and NaD. In Table~5 we present integrated $V-K$ and NaD for models with a unimodal IMF, a bimodal IMF (which diminishes the contributions of these very low-mass stars) and an IMF with a lower-mass cutoff of precisely $0.6M_{\odot}$. As one expects their 
influence is larger for steeper IMF slopes and greater ages. See also Fig.~14.

\onecolumn
\begin{table}[htb]
\scriptsize
\begin{center}
\begin{tabular}{lcccccc}
\hline
\hline
Color/Index & mag & Z=0.02, 8~Gyr & Z=0.05, 8~Gyr &
Z=0.02, 16~Gyr &  Z=0.05, 16~Gyr & max. rel. error \\
\hline
$U-V$    & y &   0.156  &   0.207  &   0.121 &    0.162 &    0.207 \\
$B-V$    & y &   0.060  &   0.085  &   0.039 &    0.060 &    0.085 \\
$V-R$    & y &   0.019  &   0.035  &   0.003 &    0.014 &    0.035 \\
$V-I$    & y &   0.030  &   0.063  &  -0.010 &    0.017 &    0.063 \\
$V-J$    & y &   0.115  &   0.122  &  -0.054 &    0.009 &    0.122 \\
$V-K$    & y &   0.163  &   0.170  &  -0.070 &    0.017 &    0.170 \\
CN$_1$   & y &   0.029  &   0.041  &   0.033 &    0.040 &    0.041 \\
CN$_2$   & y &   0.031  &   0.045  &   0.037 &    0.043 &    0.045 \\
Ca4227   & n &   0.028  &   0.125  &  -0.033 &    0.058 &    0.125 \\
G4300    & n &   0.043  &   0.010  &   0.030 &    0.004 &    0.043 \\
Fe4383   & n &   0.079  &   0.063  &   0.045 &    0.035 &    0.079 \\
Ca4455   & n &   0.071  &   0.079  &   0.038 &    0.053 &    0.079 \\
Fe4531   & n &   0.050  &   0.064  &   0.023 &    0.036 &    0.064 \\
Fe4668   & n &   0.132  &   0.113  &   0.103 &    0.085 &    0.132 \\
H$\beta$ & n &  -0.067  &  -0.088  &   0.000 &   -0.007 &    0.088 \\
Fe5015   & n &   0.067  &   0.068  &   0.051 &    0.059 &    0.068 \\
Mg$_1$   & y &   0.010  &   0.025  &   0.000 &    0.014 &    0.025 \\
Mg$_2$   & y &   0.011  &   0.027  &  -0.003 &    0.012 &    0.027 \\
Mg$_b$   & n &   0.009  &   0.023  &  -0.030 &   -0.010 &    0.030 \\
Fe5270   & n &   0.059  &   0.070  &   0.029 &    0.039 &    0.070 \\
Fe5335   & n &   0.056  &   0.073  &   0.019 &    0.041 &    0.073 \\
Fe5406   & n &   0.086  &   0.103  &   0.042 &    0.062 &    0.103 \\
Fe5709   & n &   0.099  &   0.100  &   0.088 &    0.092 &    0.100 \\
Fe5782   & n &   0.125  &   0.165  &   0.094 &    0.126 &    0.165 \\
NaD      & n &  -0.004  &   0.045  &  -0.051 &   -0.013 &    0.051 \\
TiO$_{I}$& y &   0.000  &   0.005  &  -0.002 &    0.002 &    0.005 \\
TiO$_{II}$& y &  0.001  &   0.015  &  -0.004 &    0.009 &    0.015 \\
Ca$_{II}$1& n &  0.030  &   0.065  &   0.015 &    0.053 &    0.065 \\
Ca$_{II}$2& n &  0.043  &   0.065  &   0.014 &    0.053 &    0.065 \\
Ca$_{II}$3& n &  0.062  &   0.107  &   0.026 &    0.098 &    0.107 \\
Mg$_I$   & n &   0.009  &   0.002  &  -0.033 &   -0.026 &    0.033 \\ 
\hline
\end{tabular}
\caption{Predicted differences in integrated colors and indices for SSPs with and without the inclusion of the HB+AGB phases of stellar evolution. The given numbers indicate the relative errors in magnitude, or as a fraction of the line-strength}
\end{center}
\footnotesize
\begin{center}
\begin{tabular}{l|rrr|rrr|rrr|rrr}
\hline
\hline
&\multicolumn{12}{c}{$\mu=0.35$}\\\hline
&\multicolumn{6}{c|}{$V-K$}&\multicolumn{6}{c}{NaD}\\
&\multicolumn{3}{c}{1 Gyr}&\multicolumn{3}{c|}{16 Gyr}&\multicolumn{3}{c}{1 Gyr}&\multicolumn{3}{c}{16 Gyr}\\
Z&U&B&O&U&B&O&U&B&O&U&B&O\\\hline
0.008&2.70&2.70&2.70&2.94&2.94&2.90&1.29&1.28&1.27&2.46&2.44&2.13\\
0.02 &2.97&2.97&2.97&3.30&3.30&3.27&1.97&1.97&1.95&3.56&3.54&3.26\\
0.05 &3.11&3.11&3.11&3.54&3.54&3.50&2.95&2.95&2.93&5.08&5.07&4.83\\\\
&\multicolumn{12}{c}{$\mu=1.35$}\\\hline
&\multicolumn{6}{c|}{$V-K$}&\multicolumn{6}{c}{NaD}\\
&\multicolumn{3}{c}{1 Gyr}&\multicolumn{3}{c|}{16 Gyr}&\multicolumn{3}{c}{1 Gyr}&\multicolumn{3}{c}{16 Gyr}\\
Z&U&B&O&U&B&O&U&B&O&U&B&O\\\hline
0.008&2.55&2.54&2.53&2.99&2.94&2.87&1.31&1.28&1.22&2.92&2.71&2.17\\
0.02 &2.82&2.81&2.80&3.34&3.29&3.23&1.94&1.92&1.86&4.00&3.83&3.33\\
0.05 &2.97&2.96&2.95&3.59&3.54&3.46&2.83&2.82&2.77&5.43&5.32&4.92\\\\
&\multicolumn{12}{c}{$\mu=2.35$}\\\hline
&\multicolumn{6}{c|}{$V-K$}&\multicolumn{6}{c}{NaD}\\
&\multicolumn{3}{c}{1 Gyr}&\multicolumn{3}{c|}{16 Gyr}&\multicolumn{3}{c}{1 Gyr}&\multicolumn{3}{c}{16 Gyr}\\
Z&U&B&O&U&B&O&U&B&O&U&B&O\\\hline
0.008&2.49&2.38&2.33&3.22&2.97&2.84&1.69&1.43&1.21&4.05&3.19&2.23\\
0.02 &2.76&2.65&2.60&3.56&3.32&3.20&2.25&2.04&1.83&5.05&4.31&3.43\\
0.05 &2.93&2.82&2.76&3.86&3.57&3.43&2.98&2.85&2.68&6.19&5.72&5.04\\
\hline
\end{tabular}
\caption{Test of the influence of the stars of masses lower than 0.6~M$_{\odot}$ on the integrated $V-K$ color and NaD index for two ages (1 and 16 Gyr) and three metallicities (Z=0.008, Z=0.02 and Z=0.05). 'U' means that the observables
were obtained using a unimodal IMF, which includes all the stars up to
0.0992~M$_{\odot}$, 'B' means that we used a bimodal IMF which diminishes the influence of these very low-mass stars, and 'O' means that we avoided the
low-mass stars by taking an IMF with a lower mass-cutoff of 0.6~M$_{\odot}$. 
As expected the influence of such stars is larger for steeper IMF slopes and greater ages}
\end{center}
\end{table}
\clearpage
\twocolumn

\section{The chemo-evolutionary model}
In this section we present a new and fully elaborated spectrophotometric stellar population synthesis model which, apart from an assumed IMF, needs only an analytical functional form for the SFR (see 3.1) to follow the evolution of a 
galaxy from an initial gas cloud to the present time, including the chemical evolution by taking into account mass loss from stars and supernovae. We explain the ingredients of the models together with the method of calculation in 3.2,
 while all the details about the ejecta are explained in 3.3. The output from the code will be effectively a mixture of SSPs of different ages and metallicities, which have been calculated in the way described in Section 2. 
We provide here a discussion of the influence of each of the main input 
parameters and we compare our chemical evolution model with the model of Arimoto \& Yoshii (1986) (hereafter AY86). Finally we present some tables with model 
observables obtained for a set of representative input parameters.

\subsection{The Star Formation Rate}
To quantify how much gas is converted to stars at each time, we need to define the Star Formation Rate (SFR) C(t). We assume that it is proportional to a 
power, $k$, of the fractional gas mass $f_{g}(t)$ ($f_{g}(t)=M_{g}(t)/M_{G}(t)$ where $M_{g}(t)$ is the mass of the gas at time $t$ and $M_{G}(t)$ is the total
 mass of the zone at that time, so that $M_{G}(t)=M_{g}(t)+M_{s}(t)$ where $M_{s}(t)$ is the stellar mass, including remnants. Then

\begin{equation}  
C(t)=\nu f_{g}(t)^{k}
\end{equation}

\noindent
The power $k$ is observationally estimated at between 1 and 2 according to Schmidt (1959), or $k=1.3\pm0.3$ (Kennicutt 1989). In this paper we take $k=1$.
 $\nu$ is a constant, which fixes the timescale of star formation (see Arimoto \& Yoshii, 1986,1987) and should depend on the overall physical state of the 
gas: temperature, density, and magnetic field strength etc. 

\subsection{The stellar population synthesis and the chemical evolution}
We broadly follow the mathematical formalism established by AY86 and Casuso 
(1991). We assume a fixed volume, a well-defined zone, in the galaxy under study, initially containing gas, with mass $M_{g}(t=0)$ which in the present 
paper is taken as fixed, i.e. no gas flows into or out of the volume considered. When the physical conditions appropriate for gravitational collapse are reached stars are formed, with the appropriate SFR and the IMF. 

The implementation is as follows:
during a specified time interval $\Delta t$, a generation of stars, with 
properties determined by the SFR, the IMF and the metallicity at that moment, is created if $f_{g}(t)$ is greater than a certain lower limit $f_{{g}_{min}}$. Successive generations of stars will be formed until the present time (or a 
specified epoch $T_{G}$) so that the light which is observed comes from a composite set of stars from all the generations, (SSPs), each with its 
corresponding age and metallicity, surviving at time $T_{G}$.
  
\subsubsection{The stellar population model} 
We start by assuming that all the mass is in gaseous form, i.e. zero initial mass fraction in stellar form ($M_{s}(t=0)=0$), therefore $f_{g}(t=0)=1$, and
that the gas has a certain metallicity (usually $Z(t=0)=0$). The evolution of the fractional gas mass $f_{g}(t)$ is governed by the differential equation

\begin{equation} 
{df_{g}(t) \over dt}=-C(t)+F(t)
\end{equation}

\noindent
where F(t) is the fractional mass of gas (expelled by stars and due to inflow from outside the zone) entering the zone per unit time at time t, during the last time-interval $(t,t-\Delta t)$. We first must calculate $F(t)$, which 
comprises the total ejecta (from the stars which had been created before the given epoch) in the most recent time-interval $(t,t-\Delta t)$, 
\begin{equation}
F(t)=\int_{\Delta t}^{t}\int_{m_{w}(t')}^{m_{d}(t'-\Delta t)}B(m,t-t')R_{z}(m,t')dmdt'+P(t)
\end{equation}

\noindent
where the integration over time represents the contributions to the fractional gas mass of the stars of each SSP created until $t$. A given SSP has an age of 
$t'$ at time $t$. Therefore, the last contributing SSP has an age $t'=\Delta t$ and the oldest one is the first SSP, with an age $t'=t$. In this equation:

\noindent
$B(m,t-t')$ (called the {\em birth function}) is the gas mass fraction per unit time per unit mass, which goes into stars of mass $m$ at time $t-t'$, divided by the total mass of the zone.

\noindent
$R_{z}(m,t')$ is the gas mass fraction ejected by a star of mass $m$ and metallicity Z until the time $t'$, or during its {\it lifetime} $t_{m}$ if $t' \geq t_{m}$.

\noindent
$m_{w}(t')$ is the lowest mass corresponding to a post-RGB star which is ejecting matter via winds at the age $t'$. 

\noindent
$m_{d}(t'-\Delta t)$ is the mass of the star whose {\em lifetime} $t_{m}$ is equal to the age $t'-\Delta t$.

\noindent
$P(t)$ is a term which represents the net inflow of material entering the zone.

Splitting the integral over the mass into two terms we get:
\begin{eqnarray}
F(t)&=&\int_{\Delta t}^{t} \left [ \int_{m_{w}(t')}^{m_{d}(t')}B(m,t-t')R_{z}(m,t')dm+ \right. \nonumber \\
&&\left. \int_{m_{d}(t')}^{m_{d}(t'-\Delta t)}B(m,t-t')R_{z}(m,t_{m})dm \right ] dt'\nonumber \\
+P(t)
\end{eqnarray}
\noindent
where the first term represents the contribution to the total ejecta from stars in post-RGB stages, whose matter is ejected into the interstellar medium via 
winds, and the second term represents the contribution from the stars which at time $t'$ had already reached the end of their evolution and ejected the whole of their initial masses except their remnants.  
The masses $m_{w}(t')$, $m_{d}(t')$ and $m_{d}(t'-\Delta t)$ are obtained from the corresponding isochrones. We note that these quantities are approximate, 
because they have been obtained from the closest isochrone to the required age which has the closest metallicity to that of the desired SSP. However the 
fractions $R_{z}(m,t')$ are obtained directly from the isochrones (see Section 2.1.2) and the $R_{z}(m,t_{m})$ from the final ejecta (see Section 3.3).
 
For analytical simplicity we separate B(m,t) into time-dependent and mass-dependent terms, C(t) and $\Phi(m)$, yielding
 
\begin{equation}
B(m,t)=\Phi (m)C(t)
\end{equation}

We have solved the equation (15) without using the approximation of 
instantaneous recycling (Tinsley 1980), but evaluating $f_{g}(t)$ at time $t$ using the previous values of $f_{g}(t-\Delta t)$, $C(t-\Delta t)$ and $F(t-\Delta t)$ via 
 
\begin{equation}
f_{g}(t)=f_{g}(t-\Delta t)+\frac{df_{g}(t)}{dt}\Delta t
\end{equation}

If the resulting value of $f_{g}(t)$ is greater than a pre-set value $f_{{g}_{min}}$ the SFR, $C(t)$, is evaluated from Eq.~15, but if it is smaller no stars are being formed and therefore $C(t)$ is set to 0. We apply this 
threshold based on a suggestion by Kennicutt (1989) that below a certain critical gas density there is no massive star formation at all. However,
Caldwell {\it et al.} (1994) conclude the opposite. For that reason we decided
for the present to set this free parameter $f_{{g}_{min}}$ to 0. 

\paragraph{Mass Conservation.} In order to guarantee mass conservation in our calculations, we have normalized to unity the constant $\beta$ at each time-step $\Delta t$: this is performed numerically and corrects errors generated by the
 need to discretize the time evolution. 

The finite time step also leads to the following problem. Consider a group of post-RGB stars which at time $t$ are ejecting matter into the ISM so that their contributions are included in $F(t)$. Let us now assume that at some point 
during the next step $t+\Delta t$ these stars die, so that their total contributions $R_{z}(m,t_{m})$ are included in $F(t+\Delta t)$. Therefore $F(t+\Delta t)$ also contains the part of the ejected matter previously 
computed. However at this time $t+\Delta t$ this gas, could for example go into the formation of new stars but at the same time was computed as a newly ejected matter. To correct it, we subtract this gas from $F(t+\Delta t)$.

\paragraph{Infall and Outflow.}
In this paper we consider a {\em closed-box model} in which the zone does not suffer any kind of interchange of gas with its neighborhood. However, to be more general the equations written above include this possibility by means of 
the term $P(t)$ which can have different forms (see for example Lacey $\&$ Fall (1985), or Clayton (1985,1986). Obviously, for a {\em closed-box model} the term $P(t)$ is $0$. 

\subsubsection{Chemical evolution}
Historically, see for example Tinsley (1980) and AY86, the total fraction of the heavy elements ejected by a star of initial mass $m$ was written as

\begin{equation} 
E_{z}(m)=Q_{z}(m)+Z_{g}(t-t_{m})(R_{z}(m,t)-Q_{z}(m))
\end{equation}

\noindent
where $Z_{g}$ represents an average metallicity $Z$ or that of any individual element Fe, O, etc.; $Z_{g}(t-t_{m})$ is the initial $Z$ content at the time of the star formation and $Q_{z}(m)$ is the ratio of the mass fraction of new 
metals (or this could refer to some particular element) synthesized in a star of mass $m$ and ejected, to the mass of the star. This equation was re-written by Maeder (1992) as

\begin{equation} 
E_{z}(m)=Q_{z}(m)+Z_{g}(t-t_{m})(R_{z}(m,t))
\end{equation}

\noindent
This change is due to a correct use of the definition of the yields, which are only the new and not the total fraction of metals ejected, so there is no reason to subtract $Q_{z}(m)$ a second time. Therefore the chemical evolution in our zone can be accounted for in this model as

\begin{eqnarray}  
{dZ_{g}(t) \over dt}&=&{1 \over f_{g}(t)} \int_{\Delta t}^{t}\int_{m_{d}(t')}^{m_{d}(t'-\Delta t)}B(m,t-t')[Q_{z}(m) \nonumber \\
&& + Z_{g}(t-t')R_{z}(m,t')-R_{z}(m,t')Z_{g}(t)]dmdt' \nonumber \\
&& + P(t)(Z'_{g}(t)-Z_{g}(t))
\end{eqnarray}

\noindent
where $Z'_{g}(t)$ is the metallicity of the neighborhood of the zone at time $t$, and $m_{d}(t')$ is the mass of the star whose lifetime $t_{m}$ is equal to the age of the corresponding SSP $t'$ at the time at which the computation is made.

Next the code calculates the corresponding metallicity $Z_{g}(t)$ using an equation with a similar form to Eq.~19. 

All this is done for each discrete time-step ($\Delta t$) until the desired time for the region under consideration $T_{G}$. When $t=T_{G}$ we obtain the final model results for the chemical evolution and star formation history of the zone studied. 

\paragraph{The approximation when $f_{g}(t) \rightarrow 0$.}

To avoid any numerical instabilities in the calculation of $Z(t)$ when $F(t)\Delta t \sim f_{g}(t)$ we use the following approximation for Eq.~22:
\begin{eqnarray}  
{Z_{g}(t)}&=&{1 \over F(t)} \int_{\Delta t}^{t}\int_{m_{d}(t')}^{m_{d}(t'-\Delta t)}B(m,t-t')[Q_{z}(m) \nonumber \\
&& + Z_{g}(t-t')R_{z}(m,t')]dmdt' \nonumber \\
&& + P(t)(Z'_{g}(t)-Z_{g}(t))
\end{eqnarray}
  
\begin{figure}
\caption{Comparison of the time variation of the metal content and of the fractional mass of gas computed here with those obtained by AY86 (large symbols) for different parameter sets ($\nu,\mu$), where $\nu$ is in units of $10^{-4} 
Myr^{-1}$ (the solar neighborhood value given by these authors is 1.92 in our units). Models with equal $\mu$ have the same point type. To be consistent with AY86 our results were obtained using a unimodal IMF for which $m_{l}=0.05 
M_{\odot}$ and $m_{u}=60 M_{\odot}$. For $\nu=1.92$ and $\mu=1.35$ the empirical age-metallicity relation of Twarog (1980) is well reproduced. The wiggles that 
appear when working at high SFR regimes are due to numerical instabilities which were diminished by using the approximation of Eq.~23. This is caused by the 
virtual absence of available gas. However this effect does not have any 
influence on the observed light of the galaxy, because at these high SFR most of the stars we observe at the present time were created during the first Gyr.}
\end{figure}

\subsection{The total ejecta}
As explained above, the gas-mass fraction $R_{z}(m,t)$ ejected by a star until a given time $t$, is obtained from the isochrones, but since for a star which does not end its evolution as a WD, the core C-ignition phase is the last stage 
included in the above referenced isochrones, we do not know the real mass of the remnant. This implies that we cannot calculate the total gas-mass fraction 
ejected by a star during the whole of its life $R_{z}(m,t_{m})$. We also need the stellar yields, $Q_z(m)$, to account for the chemical evolution using 
Eq.~22. Both quantities depend on the mass of the star and on its initial composition but, unfortunately, there is a lack of extensive tables giving these
 numbers, especially for non-solar metallicities. We chose to use the results of Renzini \& Voli (1981) for the intermediate stars ($1-8M_{\odot}$) and those of
 Maeder (1992) for massive stars. In these studies tables are available for solar metallicity and also for $Z=0.004$ and $Z=0.001$. 

Because the code follows the enrichment of the ISM in terms of $Z$, instead of individual elements, we have computed $Q_z(m)$ for the case of intermediate 
stars by adding all the metal contributors contained in the tables of Renzini \& Voli for the case of $\eta=1/3$ and $\alpha=0$: the parameter of the rate of mass loss in the empirical formulation by Reimers (1975) and the rate of the 
mixing-length to the pressure scale height. For all these intermediate stars we have used the remnant values only for those stars which do not end their lives as a WD because, as was pointed out above, these can be obtained directly from 
the isochrones. For massive stars we took both quantities from the tables of Maeder (1992), interpolating linearly between the values given for the two cases of mass-loss rates for stars more massive than $20M_{\odot}$ with solar 
composition. Finally, to obtain both quantities at the compositions of the published isochrones, linear interpolations were made between $Z=0.004$ and 
$Z=0.02$ to obtain the values for $Z=0.008$ in the case of intermediate stars while $Z=0.001$ and $Z=0.02$ were used to obtain the values for $Z=0.004$ and $Z=0.008$ for massive stars. To avoid extrapolation, we used values for
 $Z=0.004$ for intermediate-mass stars with lower metallicities, and values for $Z=0.001$ for massive stars of these metallicities. For the same reason, we took the solar values for higher metallicities for both kind of stars. 

All this information is kept in a file in which we interpolate linearly to obtain a more finely divided distribution of stars according to their initial mass. We must point out that we have taken into account only the fractions of
 new metals ejected $Q_z(m)$ when the stars die, and not in their previous stages. However, this is not the case for $R_{z}(m,t)$ which was computed from the isochrones. This approximation does not introduce any appreciable error,
 since the fraction of ejected gas coming from living stars during a given period period $\Delta t$ is much smaller than that of more massive dying stars and therefore the fraction of metals that is added is even less important. In
 any case, for these stars their total fraction $R_{z}(m,t_{m})$ or $Q_z(m)$ is always taken into account in the following period $\Delta t$. 
\begin{figure}
\plotone{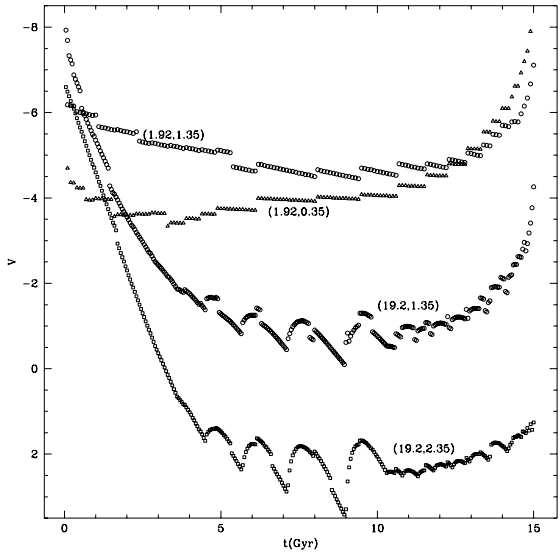}
\caption{Plot of the contributions of each constituent SSP to the integrated light in V at 15~Gyr. We used the same models as in Fig.~11. Values of the 
parameters ($\nu$,$\mu$) are shown beside the corresponding curves.}
\end{figure}
\clearpage

\subsection{The final distribution of the stars and the integrated observables}
To obtain the final distribution of the stars we scan all the created SSPs, with ages between 0 and $T_{G}$, and mark those stars whose lifetimes allow them to 
be still emitting light at the present time $T_{G}$. The number of stars N(m,t) corresponding to each SSP and which survive to the present time, with masses in 
the range ($m$,$m+\Delta m$) formed in the time interval ($t$,$t-\Delta t$) is calculated using:  

\begin{equation} 
N(m,t)={B(m,t) \over m}M_{G}(t)\Delta m\Delta t
\end{equation}

Once this distribution of stars is known we integrate the fluxes $F_{\lambda}$ of all surviving stars of all the SSPs in each photometric band 
\begin{equation}
F_{\lambda}=\int_{\Delta t}^{T_{G}}\int_{m_{l}}^{m_{t'}}N(m,t')F_{{\lambda}_{i}}(m,t')dmdt'
\end{equation}
\noindent
where $m_{t'}$ is the mass of a star which dies at $t=t'$. Finally, the code calculates the integrated set of colors.  

We proceed as in 2.4 to calculate the integrated line indices but now taking into acccount that the code yields a stellar distribution that is composed of the set of SSPs whose stars are still alive. For each surviving star of each 
created SSP, the code takes the $\log\,g$ and the $T_{eff}$ directly from the isochrones and generates the metallicity from the chemical evolution in $Z$ which is transformed to $[M/H]$, as in Eq.~10, but now $Z$ is the metallicity of the closest isochrone and not exactly the one given by the model output. A luminosity class, as defined in Eq.~9, is assigned to each contributing star.
 Then it calculates the individual contribution of a star to the desired line index via the corresponding fitting functions and integrates:
\begin{equation}
W_{i}=\frac{\int_{\Delta t}^{T_{G}}\int_{m_{l}}^{m_{t'}}W(m,t')N(m,t')F_{c}(m,t')dmdt'}{\int_{\Delta t}^{T_{G}}\int_{m_{l}}^{m_{t'}}N(m,t')F_{c}(m,t')dmdt'}
\end{equation}
\noindent
where $F_{c}(m,t')$ is the flux of the continuum of the star of mass $m$ at time $t'$ obtained by a linear interpolation as in Eq.~12.

\subsection{Input to the model}
In this section we investigate the influence of the most important input parameters; basically we focus our attention on $\nu$, $\mu$ 
and the age. These are the main parameters determining the star formation and the chemical evolution of a desired region of a galaxy, and therefore by varying them we can obtain model output that can be compared with observations.

\subsubsection{The free INPUT parameters}
The code needs the following free input parameters which will determine the evolution of the region under study:
 
\begin{itemize}
\item The SFR coefficient $\nu$ (in units of $10^{-4}Myr^{-1}$).

\item The minimum gas fraction below which no new stars can be created, $f_{{g}_{min}}$. In the present study we will leave $f_{{g}_{min}}$ at $0$.

\item The shape of the IMF: the unimodal or the bimodal IMF as were defined in Section 2.2. The slope $\mu$. The lower and upper mass-cutoff: $m_{l}$ and 
$m_{u}$. In the present study these are fixed at 0.0992 and 72 M$_{\odot}$. We shoulf poiny out that higher values for the lower mass-cutoff imply fewer low-mass stars, and by inference more high-mass stars so that the chemical
 evolution proceeds more rapidly and higher metallicities are reached. On the other hand a decrease in $m_{l}$ implies that a very important fraction of the mass is locked into low-mass stars preventing ISM enrichment, especially for 
IMF's with $\mu>1.35$. Generally, using a low IMF slope ($\mu<1.35$) higher metallicities are reached (more rapidly if using a high star formation coefficient $\nu$), due to the relative importance of massive stars.
 
\item The age $T_{G}$ of the zone and the time-step $\Delta t$. Because results do not depend strongly on $\Delta t$ we mainly use a time-step of $100~Myr^{-1}$ (see Section 3.5.3 for discussion).

\end{itemize}

\onecolumn
\begin{table}
\scriptsize
\begin{center}
\begin{tabular}{l|rrrrr|rrrrr}
\hline
\hline
Age &\multicolumn{5}{c|}{4~Gyr} &\multicolumn{5}{c}{17~Gyr}\\\hline
$\nu$($10^{-4}Myr^{-1}$)&1&   1&      20&      50&      50&       1&       1&      20&      50&      50\\
$\mu$      &      0.35&    1.35&    1.35&    1.35&    2.35&    1.35&    2.35&    1.35&    1.35&    2.35\\
\hline\hline
U-V        &   -0.1498& -0.0592& -0.0080&  0.0685&  0.0001& -0.0637& -0.0059&  0.0316&  0.1714&  0.0032\\
B-V        &   -0.0515& -0.0204& -0.0023&  0.0288& -0.0003& -0.0216& -0.0024&  0.0097&  0.0677&  0.0037\\
V-R        &   -0.0119& -0.0070& -0.0009&  0.0115& -0.0002& -0.0076& -0.0007&  0.0046&  0.0282& -0.0028\\
V-I        &   -0.0214& -0.0096&  0.0006&  0.0259& -0.0001& -0.0168& -0.0012&  0.0097&  0.0562&  0.0069\\
V-J        &   -0.0772& -0.0030&  0.0083&  0.0546&  0.0001& -0.0300& -0.0018&  0.0191&  0.0879&  0.0067\\
V-K        &   -0.1034&  0.0015&  0.0119&  0.0710&  0.0013& -0.0362& -0.0022&  0.0222&  0.1138&  0.0046\\
CN1        &    0.0029& -0.0009&  0.0096&  0.0182&  0.0009& -0.0015& -0.0003&  0.0135&  0.0390&  0.0019\\
CN2        &    0.0031& -0.0005&  0.0028&  0.0052&  0.0002& -0.0012& -0.0002&  0.0043&  0.0201&  0.0009\\
Ca4227     &    0.1473& -0.0491&  0.0487&  0.2448&  0.0035& -0.0113& -0.0039&  0.0479&  0.2651&  0.0074\\
G-band     &    0.5275&  0.2544&  0.0478&  0.3301& -0.0036& -0.0206& -0.0073&  0.0417&  0.3500&  0.0262\\
Fe4383     &    0.0882&  0.0442&  0.0563&  0.3307&  0.0017& -0.0105& -0.0088&  0.0458&  0.3305&  0.0351\\
Ca4455     &    0.0927& -0.0372&  0.0412&  0.2150&  0.0014& -0.0071& -0.0021&  0.0427&  0.2699&  0.0124\\
Fe4531     &    0.0815& -0.0225&  0.0370&  0.1764&  0.0019& -0.0047& -0.0016&  0.0378&  0.2133&  0.0096\\
Fe4668     &    0.1119& -0.0142&  0.0444&  0.2282& -0.0154& -0.0083& -0.0053&  0.0439&  0.2924& -0.0080\\
H$\beta$   &   -0.0113&  0.0054& -0.0080& -0.0698&  0.0019&  0.0101&  0.0017& -0.0179& -0.1422& -0.0117\\
Fe5015     &    0.0688& -0.0041&  0.0314&  0.1532&  0.0025& -0.0051& -0.0004&  0.0315&  0.1765&  0.0063\\
Mg$_{1}$   &    0.0055&  0.0007&  0.0025&  0.0091&  0.0001&  0.0001&  0.0000&  0.0041&  0.0164&  0.0002\\
Mg$_{2}$   &    0.0119&  0.0010&  0.0059&  0.0224&  0.0002&  0.0002& -0.0001&  0.0092&  0.0365&  0.0006\\
Mg$b$      &    0.0934&  0.0004&  0.0354&  0.1575&  0.0037& -0.0017& -0.0021&  0.0363&  0.1996&  0.0063\\
Fe5270     &    0.0653& -0.0027&  0.0403&  0.1962&  0.0033& -0.0020& -0.0013&  0.0358&  0.2109&  0.0097\\
Fe5335     &    0.0773& -0.0038&  0.0314&  0.1439&  0.0016& -0.0019& -0.0007&  0.0290&  0.1703&  0.0057\\
Fe5406     &    0.0937&  0.0069&  0.0362&  0.1734&  0.0028& -0.0001& -0.0005&  0.0347&  0.1999&  0.0075\\
Fe5709     &    0.0362&  0.0080&  0.0313&  0.1528&  0.0025&  0.0031&  0.0019&  0.0324&  0.1719&  0.0098\\
Fe5782     &    0.0646&  0.0238&  0.0477&  0.2302&  0.0121&  0.0046&  0.0000&  0.0448&  0.2485&  0.0260\\
NaD        &    0.1018&  0.0152&  0.0149&  0.0700& -0.0003&  0.0012& -0.0003&  0.0173&  0.0891& -0.0036\\
TiO$_{I}$  &    0.0067&  0.0005&  0.0006&  0.0033&  0.0001& -0.0002& -0.0001&  0.0009&  0.0050& -0.0002\\
TiO$_{II}$ &    0.0160&  0.0013&  0.0012&  0.0070&  0.0002& -0.0001& -0.0001&  0.0022&  0.0110& -0.0005\\
Ca$_{II}$1 &    0.0053&  0.0044&  0.0144&  0.0379& -0.0002&  0.0005& -0.0011&  0.0118&  0.0332&  0.0047\\
Ca$_{II}$2 &   -0.0018&  0.0092&  0.0173&  0.0504&  0.0006& -0.0023& -0.0005&  0.0145&  0.0531&  0.0039\\
Ca$_{II}$3 &    0.0003&  0.0075&  0.0162&  0.0405&  0.0006& -0.0016& -0.0005&  0.0134&  0.0510&  0.0034\\
Mg$_{I}$   &   -0.0187&  0.0132&  0.0051&  0.0562& -0.0003& -0.0084& -0.0011&  0.0103&  0.0690&  0.0062\\
\hline
\end{tabular}
\caption{The relative errors in the integrated colors and indices obtained by comparing two time-steps: $\Delta t =100$ and $50 Myr$. We selected values of ($\nu$,$\mu$) which together with the age do not allow the metallicity to be 
greater than Z=0.1. The given numbers are the difference in the color or index in magnitude, or this difference as a relative fraction (divided by the index at $\Delta t =100$), when the index is expressed in EW (following the convention of W94).}
\end{center}
\end{table}
\twocolumn

\onecolumn
\begin{table}
\hoffset-1cm
\tiny
\begin{center}
\hoffset-1cm
\begin{tabular}{l|rrr|rrr|rrr|rrr|rrr}
\hline
\hline
\multicolumn{16}{c}{Unimodal IMF ($\mu=1.35$)}\\\hline
$\nu$&&1&&&5&&&10&&&20&&&50&\\\hline
Age&4&10&17&4&10&17&4&10&17&4&10&17&4&10&17\\\hline\hline
Z$_{end}$  & 0.007& 0.019& 0.030& 0.036& 0.076& 0.084& 0.066& 0.057& 0.053& 0.075& 0.058& 0.042& 0.063& 0.100& 0.022\\
$<Z>$      & 0.004& 0.012& 0.015& 0.017& 0.025& 0.022& 0.025& 0.021& 0.020& 0.023& 0.020& 0.019& 0.019& 0.018& 0.018\\
(M/L)$_V$  &  0.92&  2.00&  3.34&  1.34&  3.90&  6.91&  1.82&  4.74&  7.49&  2.26&  5.06&  7.69&  2.43&  5.09&  7.65\\
U-V        &  0.401&  0.701&  0.855&  0.665&  1.118&  1.332&  0.851&  1.207&  1.322&  0.947&  1.206&  1.317&  0.951&  1.146&  1.237\\
B-V        &  0.421&  0.584&  0.665&  0.548&  0.811&  0.903&  0.673&  0.865&  0.905&  0.735&  0.868&  0.905&  0.745&  0.853&  0.880\\
V-R        &  0.349&  0.421&  0.462&  0.399&  0.521&  0.569&  0.452&  0.545&  0.572&  0.483&  0.549&  0.573&  0.491&  0.543&  0.563\\
V-I        &  0.809&  0.925&  1.002&  0.881&  1.097&  1.188&  0.961&  1.140&  1.191&  1.022&  1.149&  1.194&  1.035&  1.133&  1.172\\
V-J        &  1.593&  1.959&  2.073&  1.951&  2.213&  2.273&  2.027&  2.218&  2.243&  2.091&  2.211&  2.239&  2.079&  2.164&  2.190\\
V-K        &  2.415&  2.876&  2.993&  2.886&  3.165&  3.200&  2.968&  3.147&  3.152&  3.027&  3.132&  3.144&  3.000&  3.071&  3.086\\
CN1        & -0.153& -0.104& -0.082& -0.116& -0.031& -0.009& -0.073& -0.018& -0.015& -0.050& -0.018& -0.015& -0.059& -0.037& -0.039\\
CN2        & -0.084& -0.047& -0.029& -0.056&  0.011&  0.031& -0.022&  0.024&  0.028& -0.004&  0.024&  0.029& -0.003&  0.018&  0.019\\
Ca4227     &  0.240&  0.584&  0.775&  0.514&  1.077&  1.333&  0.737&  1.158&  1.300&  0.858&  1.171&  1.306&  0.798&  1.061&  1.177\\
G-band     & -0.184&  1.630&  2.351&  1.076&  3.694&  4.383&  2.307&  4.159&  4.294&  2.991&  4.184&  4.294&  2.812&  3.878&  3.898\\
Fe4383     &  0.037&  1.598&  2.409&  1.316&  3.681&  4.417&  2.359&  3.925&  4.238&  2.850&  3.915&  4.235&  2.597&  3.510&  3.793\\
Ca4455     &  0.289&  0.728&  0.927&  0.677&  1.258&  1.418&  0.959&  1.292&  1.359&  1.065&  1.286&  1.357&  0.984&  1.172&  1.230\\
Fe4531     &  1.126&  1.909&  2.249&  1.776&  2.780&  3.092&  2.244&  2.870&  3.007&  2.455&  2.872&  3.010&  2.311&  2.670&  2.777\\
Fe4668     &  0.145&  1.579&  2.344&  1.615&  3.687&  4.055&  2.720&  3.622&  3.744&  3.024&  3.537&  3.716&  2.723&  3.108&  3.308\\
H$\beta$   &  4.977&  3.813&  3.308&  4.320&  2.474&  1.903&  3.425&  2.061&  1.839&  2.912&  2.007&  1.818&  2.853&  2.054&  1.925\\
Fe5015     &  2.257&  3.423&  3.909&  3.339&  4.636&  4.849&  4.004&  4.609&  4.657&  4.230&  4.569&  4.642&  3.976&  4.262&  4.324\\
Mg$_{1}$   &  0.018&  0.038&  0.056&  0.035&  0.076&  0.098&  0.050&  0.080&  0.095&  0.058&  0.081&  0.096&  0.055&  0.073&  0.086\\
Mg$_{2}$   &  0.070&  0.117&  0.151&  0.112&  0.189&  0.223&  0.141&  0.194&  0.216&  0.155&  0.195&  0.216&  0.147&  0.178&  0.197\\
Mg$b$      &  1.307&  1.919&  2.420&  1.818&  2.966&  3.408&  2.254&  3.054&  3.306&  2.507&  3.059&  3.304&  2.402&  2.823&  3.050\\
Fe5270     &  0.971&  1.744&  2.075&  1.674&  2.481&  2.707&  2.037&  2.487&  2.612&  2.160&  2.478&  2.611&  2.011&  2.287&  2.410\\
Fe5335     &  0.760&  1.436&  1.780&  1.397&  2.245&  2.465&  1.817&  2.248&  2.375&  1.944&  2.234&  2.373&  1.829&  2.072&  2.209\\
Fe5406     &  0.415&  0.874&  1.111&  0.840&  1.412&  1.578&  1.106&  1.420&  1.518&  1.198&  1.414&  1.518&  1.126&  1.297&  1.399\\
Fe5709     &  0.384&  0.637&  0.733&  0.642&  0.859&  0.884&  0.757&  0.836&  0.843&  0.780&  0.825&  0.839&  0.732&  0.767&  0.782\\
Fe5782     &  0.265&  0.502&  0.596&  0.506&  0.708&  0.747&  0.601&  0.687&  0.710&  0.621&  0.680&  0.708&  0.567&  0.615&  0.645\\
NaD        &  1.289&  1.827&  2.306&  1.789&  2.824&  3.315&  2.228&  2.892&  3.268&  2.396&  2.910&  3.287&  2.338&  2.777&  3.132\\
TiO$_{I}$  &  0.032&  0.032&  0.034&  0.031&  0.038&  0.042&  0.032&  0.039&  0.042&  0.035&  0.040&  0.042&  0.034&  0.038&  0.040\\
TiO$_{II}$ &  0.039&  0.040&  0.047&  0.037&  0.058&  0.068&  0.043&  0.061&  0.067&  0.050&  0.062&  0.067&  0.048&  0.057&  0.063\\
Ca$_{II}$1 &  1.110&  1.519&  1.623&  1.587&  1.653&  1.648&  1.597&  1.603&  1.608&  1.588&  1.587&  1.603&  1.536&  1.549&  1.574\\
Ca$_{II}$2 &  3.137&  3.998&  4.212&  4.160&  4.208&  4.135&  4.124&  4.062&  4.020&  4.088&  4.016&  4.002&  3.938&  3.895&  3.900\\
Ca$_{II}$3 &  2.703&  3.496&  3.642&  3.655&  3.606&  3.469&  3.607&  3.450&  3.361&  3.545&  3.401&  3.341&  3.415&  3.305&  3.261\\
Mg$_{I}$   &  0.421&  0.617&  0.697&  0.625&  0.787&  0.829&  0.696&  0.775&  0.811&  0.714&  0.769&  0.809&  0.690&  0.740&  0.781\\
\hline
\multicolumn{16}{c}{Bimodal IMF ($\mu=1.35$)}\\\hline
$\nu$&&1&&&5&&&10&&&20&&&50&\\\hline
Age&4&10&17&4&10&17&4&10&17&4&10&17&4&10&17\\\hline\hline
Z$_{end}$  & 0.011& 0.025& 0.042& 0.049& 0.102& 0.121& 0.088& 0.094& 0.087& 0.106& 0.088& 0.068& 0.091& 0.134& 0.042\\
$<Z>$      & 0.006& 0.015& 0.028& 0.028& 0.032& 0.029& 0.031& 0.028& 0.027& 0.029& 0.027& 0.026& 0.027& 0.025& 0.025\\
(M/L)$_V$  &  0.51&  1.16&  2.00&  0.85&  2.19&  3.81&  1.08&  2.71&  4.26&  1.31&  2.90&  4.36&  1.42&  2.91&  4.35\\
U-V        &  0.448&  0.712&  0.950&  0.712&  1.095&  1.327&  0.839&  1.217&  1.375&  0.935&  1.220&  1.363&  0.957&  1.160&  1.262\\
B-V        &  0.432&  0.586&  0.711&  0.582&  0.793&  0.895&  0.661&  0.858&  0.921&  0.723&  0.865&  0.919&  0.739&  0.848&  0.882\\
V-R        &  0.349&  0.420&  0.475&  0.409&  0.513&  0.564&  0.446&  0.541&  0.576&  0.476&  0.547&  0.576&  0.486&  0.539&  0.562\\
V-I        &  0.809&  0.924&  1.008&  0.879&  1.076&  1.169&  0.946&  1.127&  1.189&  1.007&  1.139&  1.191&  1.023&  1.121&  1.164\\
V-J        &  1.658&  1.999&  2.124&  1.961&  2.236&  2.292&  2.073&  2.258&  2.283&  2.131&  2.246&  2.267&  2.139&  2.209&  2.229\\
V-K        &  2.485&  2.921&  3.076&  2.909&  3.206&  3.232&  3.042&  3.208&  3.206&  3.092&  3.183&  3.180&  3.088&  3.130&  3.128\\
CN1        & -0.146& -0.100& -0.065& -0.100& -0.034& -0.004& -0.076& -0.016& -0.001& -0.053& -0.013& -0.001& -0.054& -0.031& -0.029\\
CN2        & -0.079& -0.043& -0.015& -0.043&  0.012&  0.039& -0.022&  0.030&  0.046& -0.003&  0.033&  0.046&  0.004&  0.027&  0.032\\
Ca4227     &  0.290&  0.617&  0.901&  0.621&  1.075&  1.352&  0.757&  1.201&  1.397&  0.888&  1.230&  1.403&  0.847&  1.114&  1.246\\
G-band     &  0.185&  1.705&  2.708&  1.518&  3.486&  4.291&  2.256&  4.103&  4.486&  2.983&  4.217&  4.516&  2.817&  3.828&  3.864\\
Fe4383     &  0.337&  1.826&  3.040&  1.832&  3.775&  4.648&  2.542&  4.230&  4.780&  3.082&  4.265&  4.758&  2.900&  3.840&  4.143\\
Ca4455     &  0.379&  0.787&  1.113&  0.849&  1.293&  1.485&  1.013&  1.382&  1.501&  1.135&  1.387&  1.493&  1.067&  1.260&  1.327\\
Fe4531     &  1.297&  1.992&  2.491&  2.017&  2.800&  3.167&  2.297&  2.978&  3.209&  2.536&  3.004&  3.208&  2.410&  2.780&  2.905\\
Fe4668     &  0.388&  1.947&  3.329&  2.425&  4.144&  4.694&  3.152&  4.344&  4.668&  3.521&  4.261&  4.568&  3.326&  3.803&  4.079\\
H$\beta$   &  4.852&  3.830&  3.190&  4.105&  2.681&  2.049&  3.554&  2.215&  1.852&  3.029&  2.107&  1.812&  2.925&  2.151&  1.991\\
Fe5015     &  2.528&  3.627&  4.378&  3.783&  4.809&  5.107&  4.199&  4.928&  5.074&  4.467&  4.908&  5.036&  4.216&  4.532&  4.649\\
Mg$_{1}$   &  0.020&  0.043&  0.068&  0.043&  0.081&  0.104&  0.055&  0.089&  0.107&  0.063&  0.090&  0.107&  0.063&  0.083&  0.097\\
Mg$_{2}$   &  0.077&  0.127&  0.174&  0.128&  0.198&  0.235&  0.150&  0.211&  0.238&  0.166&  0.212&  0.237&  0.161&  0.194&  0.215\\
Mg$b$      &  1.404&  2.066&  2.704&  2.016&  3.064&  3.556&  2.357&  3.263&  3.594&  2.635&  3.287&  3.579&  2.563&  3.032&  3.283\\
Fe5270     &  1.141&  1.866&  2.348&  1.923&  2.576&  2.844&  2.150&  2.661&  2.848&  2.295&  2.661&  2.835&  2.160&  2.454&  2.595\\
Fe5335     &  0.875&  1.569&  2.115&  1.694&  2.378&  2.643&  1.957&  2.462&  2.647&  2.102&  2.451&  2.627&  2.028&  2.290&  2.446\\
Fe5406     &  0.501&  0.965&  1.317&  1.017&  1.499&  1.696&  1.195&  1.564&  1.703&  1.301&  1.560&  1.691&  1.255&  1.443&  1.562\\
Fe5709     &  0.464&  0.695&  0.855&  0.746&  0.926&  0.970&  0.821&  0.933&  0.956&  0.852&  0.922&  0.946&  0.811&  0.859&  0.883\\
Fe5782     &  0.323&  0.554&  0.699&  0.595&  0.762&  0.817&  0.655&  0.768&  0.807&  0.682&  0.760&  0.799&  0.636&  0.692&  0.728\\
NaD        &  1.278&  1.943&  2.607&  2.047&  2.952&  3.453&  2.375&  3.100&  3.514&  2.543&  3.108&  3.501&  2.549&  2.998&  3.383\\
TiO$_{I}$  &  0.029&  0.031&  0.033&  0.029&  0.037&  0.041&  0.031&  0.039&  0.042&  0.034&  0.040&  0.042&  0.033&  0.036&  0.039\\
TiO$_{II}$ &  0.032&  0.039&  0.047&  0.037&  0.057&  0.067&  0.043&  0.061&  0.068&  0.049&  0.062&  0.068&  0.047&  0.056&  0.062\\
Ca$_{II}$1 &  1.151&  1.604&  1.689&  1.644&  1.718&  1.702&  1.689&  1.686&  1.679&  1.668&  1.654&  1.657&  1.622&  1.612&  1.626\\
Ca$_{II}$2 &  3.244&  4.204&  4.339&  4.271&  4.353&  4.256&  4.337&  4.252&  4.179&  4.274&  4.174&  4.126&  4.133&  4.041&  4.016\\
Ca$_{II}$3 &  2.760&  3.669&  3.772&  3.773&  3.747&  3.580&  3.812&  3.618&  3.489&  3.723&  3.539&  3.443&  3.601&  3.433&  3.354\\
Mg$_{I}$   &  0.449&  0.643&  0.765&  0.681&  0.820&  0.854&  0.746&  0.818&  0.844&  0.760&  0.809&  0.837&  0.745&  0.780&  0.816\\
\hline
\end{tabular}
\caption{The model observables obtained with our full chemical evolutionary model for a constant unimodal and bimodal IMF with slope 1.35. The SFR 
coefficient $\nu$ is in units of $10^{-4}Myr^{-1}$, while the age is in Gyr. Z$_{end}$ indicates the metallicity obtained at the assumed age while $<Z>$ indicates the average metallicity}
\end{center}
\end{table}
\twocolumn
    
\subsubsection{Effects of changing $\mu$ and $\nu$}
The main parameters determining the evolution of the metallicity in a model are the coefficient $\nu$ of the SFR and the shape of the IMF, mainly its slope 
$\mu$. These two as well as the assumed age for the zone under study will determine the final metallicity and the average value. 

We have compared our evolution of the metallicity and the fractional gas mass with AY86, using the same IMF (with the same values of the lower and upper mass limits), a linear dependence of the SFR on the fractional gas mass, and without
 using any star formation threshold. The evolution of the metallicity of the gas, and the gas fraction are plotted in Fig.~11 for different pairs of model parameters ($\nu$,$\mu$). Despite the use of updated isochrones and stellar yields, the latter for the case of massive stars, the agreement is very good.
 Here also the empirical relation between age and metallicity of nearby stars (Twarog 1980) is fitted for $\nu=1.92$ (the solar neighborhood value given in AY86 but in our units of $10^{-4} Myr^{-1}$) and a Salpeter IMF ($\mu=1.35$)
. Despite the similarity of the chemical evolution, their resulting colors are different from ours. This is caused by the use of different isochrones and different isochrone-color conversions.
This point was already discussed in Bressan {\it et al.} (1994), who used the 
same set of isochrones we used, but different conversions.  

This figure can be used to see the effect of varying the SFR coefficient together with the IMF slope. When using a low $\nu$ only a small fraction of the available gas at each time $t$ will be used to form new stars, and therefore the
 ISM enrichment is very slow, especially for a high $\mu$, which favors low mass stars which do not eject large amounts of matter. In Fig.~12 where we have represented the contribution of each SSP to the presently observed light in the V band, we see how most of the observed light comes from the recent SSPs. 
However, for high SFR models the amount of available gas rapidly decreases (see the case of $\nu=19.2$ in Fig.~11), especially when using high IMF slopes, causing the matter to be locked up in low-mass stars. In that case the observed light comes mostly from the oldest SSPs. For high SFR the metallicity usually 
increases much faster, but for $\nu\geq 10$, at a certain time the metallicity reaches a maximum and starts to decline due to ejection of unprocessed matter from the oldest and numerous low-metallic SSPs stars. Obviously, here one should understand that the highest resulting metallicity is reached when combining 
strong $\nu$'s with low $\mu$'s. This effect can also be seen when combining a normal $\nu=1.92$ with $\mu=0.35$ obtaining $Z>0.1$ values (see Fig.~11). At $Z=0.1$ the models are still not good enough to be compared with observations.
  
Following Arimoto \& Yoshii (1987), a typical time\-scale value for elliptical galaxies is in around $\sim100\times 10^{-4} Myr^{-1}$. However, to fit different kinds of objects, often other values over a wide range ($0.1-200$) are needed. For the same reason in practice, a considerable range in $\mu$ around 
1.35 (Salpeter) might be needed in order to fit the observations.

\subsubsection{The time-step} 
$\Delta t$ is the time-step used in the calculation. In principle this parameter should be infinitely small but one has to find a value sufficiently large that the model does not spend too much computing time, and sufficiently small that 
the results remain accurate. In theory, this parameter has to be smaller than the cooling time of a cloud to form a new SSP. Empirically we have made tests to ascertain for which $\Delta t$ we can still obtain reliable results. Table~6
shows the dependence of the final results on this parameter for a reasonable range of values (we selected values of $\nu$ and $\mu$ which at the assumed age do not allow the metallicity to be greater than Z=0.1, see Fig.~11). In general the relative error for the colors and indices is smaller than a few percent, 
except for very high SFR coefficients, when, for $\nu=50$, this error is around $\sim 20\%$. The worst colors again are U-V and $V-K$. This comparison shows that 
for most cases, a value of $100~Myr.$ is reasonable and is sufficient to obtain reasonable results. However, when working with high $\nu$ and/or low $\mu$ it is advisable to use shorter time-steps. Obviously, the use of much shorter $\Delta 
t$ is more important for the study of systems with recent starbursts, but is not necessary for old systems. In fact, a useful compromise can be to use shorter time-steps for the last $100~Myr.$ time interval only.

\subsubsection{The static option}
If we remove all the evolutionary aspects of our code we obtain a model consisting of a unique SSP as in Section 2. This option could in fact be 
obtained by equating the time-step $\Delta t$ to the desired age of the galaxy, and by using the initial metallicity throughout.

\subsection{The observables from our closed-box evolutionary model}
Here we assumes that our region does not interchange any matter with the neighborhood, the so-called closed-box model. In Table~7 we present the observables and metallicities (the final and the average values) obtained with 
unimodal and bimodal IMFs constant in time, with slope $\mu$=1.35 for different SFR coefficients and ages. For the sake of brevity and since we want to apply our model to early-type galaxies we present in this Table results obtained only 
for $\mu$=1.35 because with this value we obtain metallicities which are not too far from solar, as can be seen in Fig.~11. As expected, the bimodal IMF always leads to higher metallicities since the weight of low-mass stars is diminished 
and therefore the chemical evolution is more influenced by massive stars. However, as explained above, for real galaxies one might need a range in $\mu$. This Table and others with further data can be obtained from the authors.

\section{Application of our model to three early-type galaxies}
In this section we apply the model we described in the previous sections to the central regions of a few standard early-type galaxies. We first apply the {\em 
static} version of our models, and later the evolutionary version. Although {\em static} studies have been carried out before (e.g. Peletier 1989, Worthey {\it et al.} 1992, etc.) in this paper and its companion (Paper II) many more spectral line indices are analyzed than is generally done. The evolutionary 
aspect of our model is completely new. Up to now, nobody has tried to fit at the same time colors and absorption line strengths of elliptical galaxies with an
 evolutionary model. For that purpose we have obtained high quality spectroscopic observational data of two standard giant ellipticals (NGC 3379 and NGC 4472) and the bulge of the Sombrero Galaxy (NGC 4594). The details of these 
observations (including the conversion of the line-strengths to the Lick system) are to be found in Paper II. Here as a sample we will analyze the values for the three 
galaxies at 5 arcsec from the center (to avoid problems related to the seeing): along the minor axis for the Sombrero galaxy and along the major axis for the 
giant ellipticals. The colors for the two giant ellipticals are taken from 
Peletier {\it et al.}(1990a) and Peletier {\it et al.}(1990b) while those for
 the Sombrero galaxy are taken from Hes \& Peletier (1993). In this paper we will concentrate on what we can learn about the models, while in Paper II the 
observations are discussed and emphasis is put on what we can learn about the 
stellar content of elliptical galaxies.

\begin{figure}
\plotone{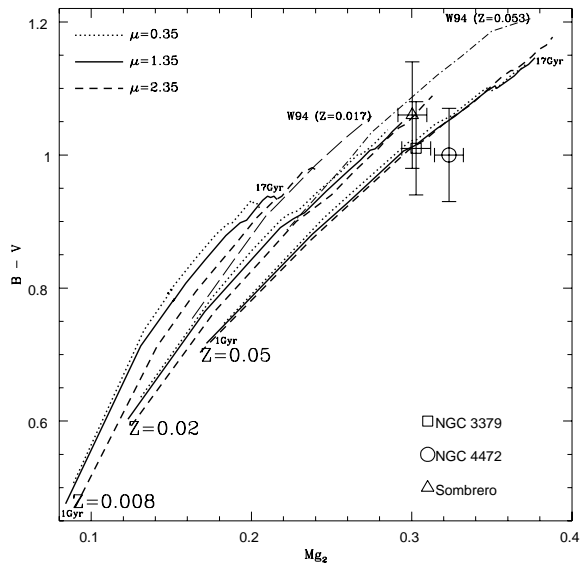}
\caption{Synthetic diagram of $B-V$ vs. Mg$_{2}$ obtained by calculating
single-age stellar populations (SSP) for different ages (ages range from 1 to 
17 Gyr. along each curve shown) and metallicities, for unimodal IMFs with 
slopes: 0.35, 1.35 (Salpeter case) and 2.35. Also indicated are the observed values for the three galaxies at 5 arcsec from their centers. Here it is 
possible to see that a higher than (or at least equal to) solar metallicity is required to fit real data, whatever the age or the IMF slope. This confirms the results given in W94 for a Salpeter IMF.}
\end{figure}

\begin{figure}
\plotone{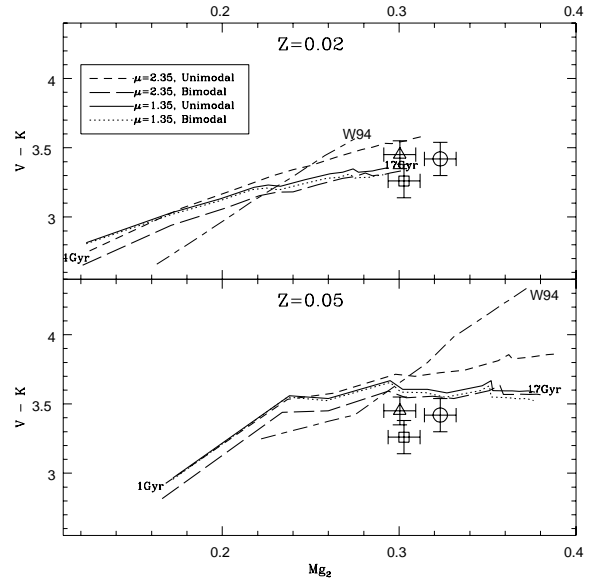}
\caption{Shown are various SSP models together with observations for our three galaxies. Also shown are models from W94 for comparison.}
\end{figure}

\begin{table}
\footnotesize
\begin{center}
\begin{tabular}{c|cccc} 
\hline
\hline
$t_{0}(Gyr.)$& 0.2 & 0.5 &  1 & 2 \\ \hline
$Z(t_{0})$ &&($\nu$,$\mu_{0}$)&&  \\ \hline \hline
$\sim 2.5 \times Z_{\odot}$&(30,0.8)&(15,0.8) &(10,1)   &(5,1)   \\
                           &(30,1)  &(15,1)   &(7.5,0.8)&(2.5,0.5)\\
                           &(20,0.5)&(10,0.5) &(7.5,1)  &(1,-1)  \\
                           &(20,0.8)&(10,0.8) &(5,0.5)  &\\
                           &(15,0)  &(7.5,0)  &(5,0.8)  &\\
                           &(15,0.5)&(7.5,0.5)&(2.5,-1) &\\
                           &(10,-1) &(5,-1)   &(2.5,0)  &\\
                           &(10,0)  &(5,0)    &         &\\ \hline

$\sim Z_{\odot}$           &(15,1)  &(7.5,1)  &(5,1)   &(1,0.5)  \\
                           &(10,0.5)&(5,0.8)  &(2.5,0.5)&(1,0.8)  \\
                           &(10,0.8)&(2.5,0)  &(2.5,0.8)&(0.5,-1) \\
                           &(7.5,0) &(2.5,0.5)&(1,0)  &(0.5,0)  \\
                           &(7.5,0.5)&        &(1,-1)  &  \\
                           &(5,-1)  &&&\\
                           &(5,0)  &&&\\
\hline
\end{tabular}
\caption{Sets of models with parameters ($\nu$,$\mu_{0}$) which drive the chemical evolution to reach values $\sim 2.5$ times solar, and solar metallicity at times, $t_{0}$, of 0.2, 0.5, 1 and 2 Gyr.}
\end{center}
\end{table} 

\subsection{Applying the single-age stellar population model}
In Figs.~13 and 14 we have plotted the models, together with our three galaxies, in two important diagnostic diagrams: $B-V$ vs. Mg$_{2}$, and of $V-K$ vs. 
Mg$_{2}$. In the first place, one sees that for each galaxy there are models which fit the data. In the two diagrams our models seem to fit better than, for example, those by W94. Looking at the two diagrams one can also see that solar 
metallicities or larger values are required to fit the observational data for the three galaxies, whatever their ages or their IMF slopes. This result was 
inferred by ourselves (Casuso {\it et al.} 1996) from an analysis of the Mg$_{2}$ index alone.

Comparing models by various authors (see Section 2.6) we have seen that $V-K$ is a very difficult color to model, due to uncertainties in the color-temperature
and color-color transformations for low-mass stars and in our knowledge of
advanced stages of stellar evolution. On the other hand it is also a very
sensitive color, due to its large wavelength baseline. In Fig.~14 we see that
$V-K$ depends considerably on the IMF slope $\mu$. We consider IMF slopes of
1.35 and 2.35 and the two shapes of IMF proposed in Section 2.2.
In this plot we see that low-mass stars are important in determining $V-K$, since there is a substantial difference between the predictions for the unimodal
and bimodal IMFs, especially for high IMFs slopes and greater ages. Therefore
the stars with masses below $0.6M_{\odot}$ have a non negligible effect, and can
make the $V-K$ color redder by $\sim0.4~mag$ for old populations (see for
details Section 2.7.1). 

Finally we see that if one includes a small amount of extinction, which reddens the colors but does not change the line indices, the quality of the fits 
degrades. In Paper II we will discuss the fit to the other colors and absorption lines in more detail.

\subsection{Applying the chemo-evolutionary model}
In an exploratory attempt to find an acceptable evolutionary fit to our galaxies we will focus our attention on various very important observables: the colors 
$V-K$ and $U-V$, and the Mg$_{2}$ and H$\beta$ indices. We first run models varying the two main parameters ($\mu$,$\nu$) and also the ages assigned to the 
galaxies. Here we also study the two forms of the IMF: the unimodal and the bimodal cases. In all these cases we use an IMF slope constant in time in a {\em closed-box} approximation. Figs.~15, 16 and 17 show the model output for the 
selected observables when we sweep our parameter space. In these plots one can 
easily see how our evolutionary scheme of stellar population synthesis cannot fit the real data. In particular we see that the Mg$_{2}$ index of the models is always too low for given colors. There is an exception when using extremely 
steep unimodal IMFs for which we can obtain $Mg_{2}\geq0.28$ (independently of the SFR that we use) but which cannot increase as rapidly as the $V-K$ color, due to the coolest very low MS stars. 

\begin{figure}
\plotone{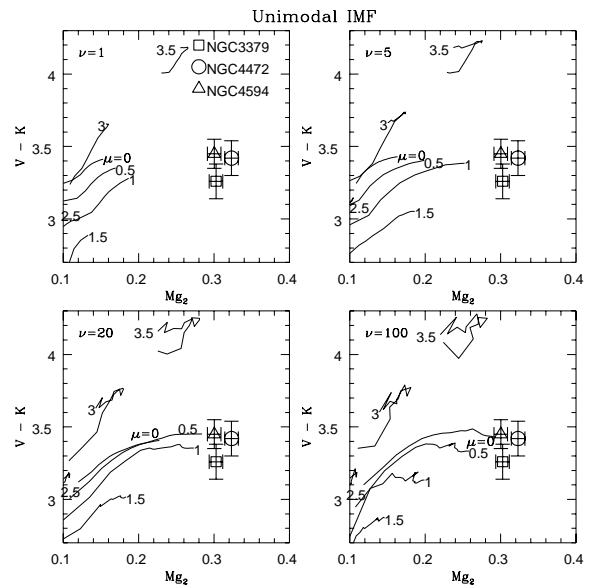}
\caption{Synthetic Mg$_{2}$-($V-K$) diagram for four SFR regimes ($\nu$=1, 5, 20 and 100$\times10^{-4}Myr^{-1}$), for a unimodal IMF with different slopes, from 0 to 4 in steps of 0.5. The age for the galaxies was varied from 1 to 17~Gyr,
 which corresponds to the last point on each curve (where the IMF slope is annotated). Observations for the three galaxies are shown.}
\end{figure}

\begin{figure}
\plotone{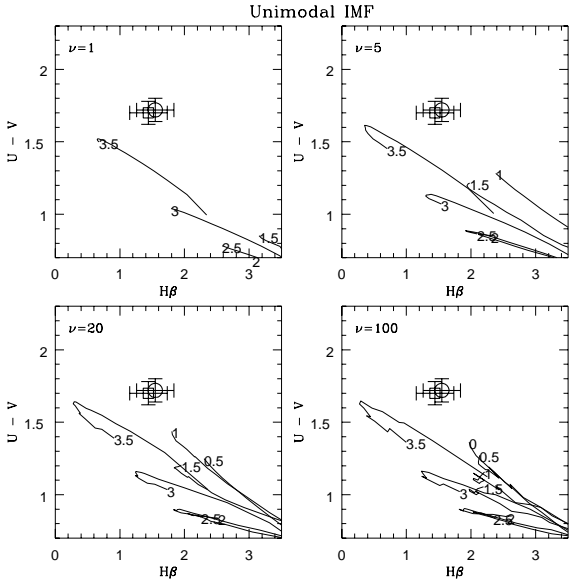}
\caption{The H$\beta$-(U-V) diagram corresponding to the same set of parameters as in Fig.~15.}
\end{figure}

\begin{figure}
\plotone{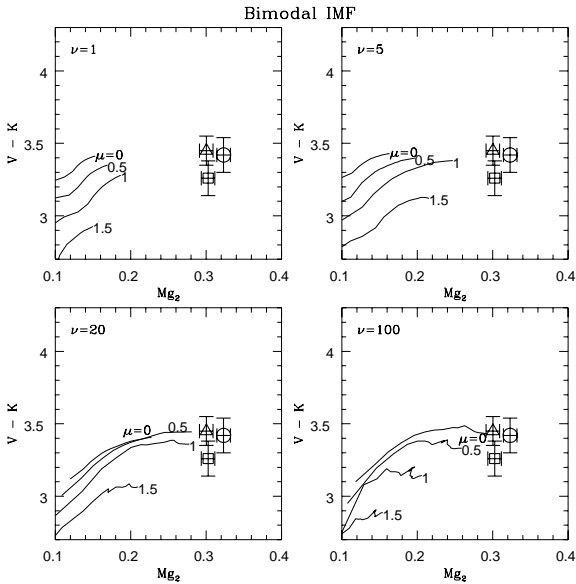}
\caption{The ($V-K$)-Mg$_{2}$ diagram obtained for the case of a bimodal IMF.}
\end{figure}

It is didactically instructive to compare the results of our models with the 
distribution of stars as a function of the metallicity with those resulting from analytical models. In Fig.~18 we show the contributions at the present time of
 the stars as a function of metallicity and age for some of the best fitting (although not acceptable) models, inferred from Figs.~15, 16 and 17. We 
have also plotted some of the so-called {\it simple analytical models} which 
assume a closed system, initially all metal-free gas and a constant IMF (see the review of Audouze \& Tinsley 1976). We see that both, the simple models and ours, yield distributions which contain important fractions of low-metallicity stars, 
explaining the mismatch in the $V-K$ vs. Mg$_{2}$. We are clearly dealing with a version of the G-dwarf problem in which we have to account for the absence of 
low metallicity stars. This is why, with our invariant IMF models, we cannot 
obtain higher values for Mg$_{2}$. If we choose a low value for $\nu$ to decrease the importance of the first SSPs, then we get more or less continuous 
star formation, and the youngest SSPs will be the brightest, and since they are
 young, their Mg$_{2}$ will be rather low, even if they are metal-rich. All this is thoroughly discussed in Casuso {\it et al.} (1996) in which we focused our 
attention exclusively on the Mg$_{2}$ index. If one looks at the results 
presented in AY86, one sees that the authors have the same problem, and that they can barely find a solution that fits a typical $B-V$ and $V-K$ for a galaxy. They 
would not have been able to fit its Mg$_{2}$ index with their model. The 
previous conclusions are more solid if we look at the  H$\beta$ vs. $U-V$ diagram, which cannot be fitted either. Finally, in Figs.~15 and 17 there are a
few models, with very low IMf slopes, which almost fit the $V-K$ vs. Mg$_{2}$ 
diagrams of this set of galaxies. These models however do not fit the corresponding H$\beta$ vs. $U-V$ diagrams (see Fig.~16).
\begin{figure}
\vskip-1.5cm
\plotone{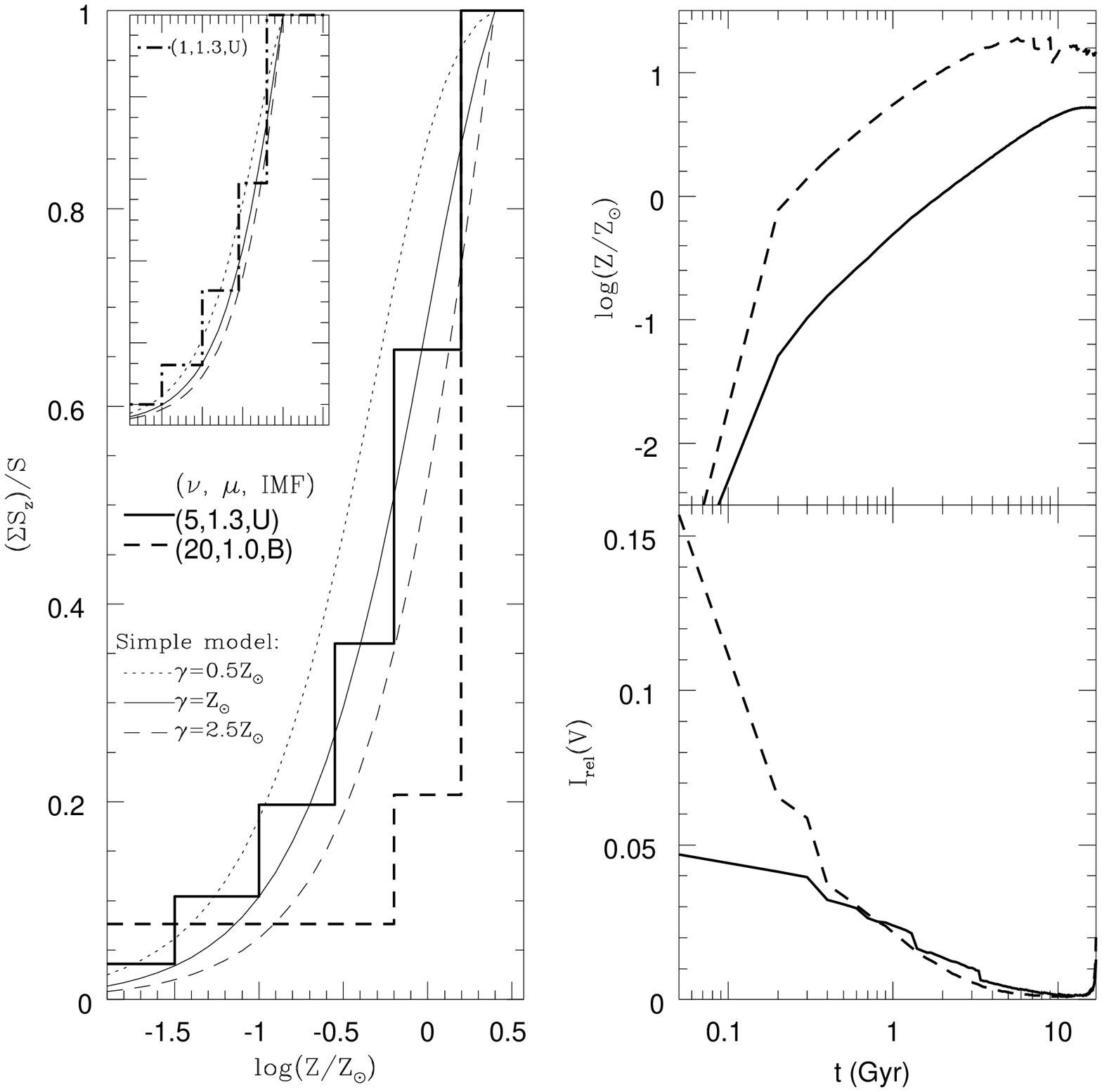}
\vskip-0.5cm
\caption{Histogram of the cumulative mass fraction of living stars as a function of the metallicity for two models, with constant IMF, 
which give fits that are the best, but not acceptable, compared to real 
data. 'U' means unimodal IMF while 'B' means bimodal. These models were selected by looking at Figs.~15, 16 and 17. We have also plotted a couple of simple 
analytical models fixing the final metallicity to match ours, but varying the
 {\em net yield} $\gamma$. To be compared with these analytical models, in the embedded figure (same scale) we plot one of our models that give rise to 
chemical evolution which yields to solar metallicity. Notice that the analytical
models almost match ours. All our models assume an age of 17~Gyr. In the second
figure we plot the corresponding chemical evolution for these two models.
Finally, in the last figure we show I$_{rel}$(V), the fraction (not cumulative)
of the total luminosity per time step in V, at the present time, of the SSPs
formed at different epochs. Notice that the cumulative fraction of stars with
metallicities lower than solar is in the range $8-36\%$. Also notice that these
stars contribute substantially to the present light, as indicated in the last
plot.}
\end{figure}

The fact that the metallicity distribution of our model agrees with the predictions of a simple analytic model does not mean however that the latter are adequate for handling in detail the variations which have in fact occurred in 
stellar populations, even in less complex systems such as ellipticals. The present model is much more flexible in including the evolutionary histories of
stars (such as mass loss) and allows us to understand better the star formation history. The fact that the stellar frequency distribution with metallicity can 
be reproduced with an analytical model does however give some support to our procedure.

\subsubsection{A variable IMF scenario}
The key question is now how to build a scenario that produces a dominant old but metal-rich population in our observed galaxies. To tackle this problem, in this
 paper we introduce a time dependence in the IMF slope as one of the ways which 
are able to yield a metallicity distribution which is skewed toward higher
 metallicities. Another would be to include some gaseous infall, but we considered this to be less lively for ellipticals. In the beginning, for a period of time ($t_{0}$) of $\sim 1~Gyr$, or even less, we favor the production of massive stars (using an IMF with low slope $\mu_{0}$), while later mainly 
low-mass stars are produced (this IMF slope is termed $\mu$). The first to propose this variable IMF scenario was Schmidt (1963). It also was analyzed by Arnaud {\it et al.} (1992), Worthey {\it et al.} (1992) and more recently by 
Elbaz {\it et al.} (1995) among others. Some authors tackled the problem by keeping the IMF slope constant (with the Salpeter index), but changing the 
lower-mass cutoff. In our scenario, at the beginning short-lived massive stars 
quickly enrich the ISM but these stars, at the same time, do not contribute to
 the light we are observing now. This light comes essentially from stars created later from metal-rich gas and a steeper IMF. 

This scenario is plausible physically because:
\begin{itemize}
\item It is likely that the central density and temperature were initially,
high, which favored high-mass star formation (see starbursts references) in
these massive galaxies, so that the IMF slope could well have been smaller 
than 1.

\item Later, conditions tend towards those of the solar neighborhood, so that $\mu$ becomes larger.
\end{itemize}

If we do not want to include a specific treatment for the galactic equilibrium between SN rates and the galactic mass-dependent binding energy to predict when
 gas densities are able to form stars, we have to introduce a phenomenological 
parameter, which is the duration of this short period in which massive stars
 were favored. For this purpose we first plot in Fig.~19 the chemical evolution during the first 2 Gyr. for flatter IMF slopes ($-1\leq\mu_{0}\leq1$) for 
different SFR regimes (characterized by the $\nu$ coefficients). The results 
here do not depend very much on whether we choose a unimodal or bimodal IMF (in the latter case the metallicities that are reached are slightly higher). Because
 we are limited by the isochrones, which are available only for solar and 2.5 
times solar metallicities, we will analyze those chemical evolution parameters 
for which the two metallicities are reached in different periods of time: 0.2, 0.5 1 and 2~Gyr. Table~8 summarizes these results. The main difference for two 
pairs of chemical evolution models which reach the same metallicity level at this initial period is that for lower $\mu_{0}$ the contribution of the earliest SSPs to the present light is less important. 
\begin{figure}
\plotone{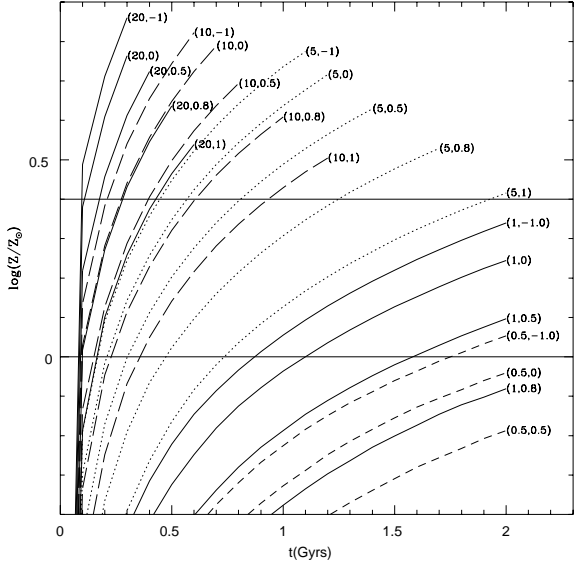}
\caption{The chemical evolution in the first 2~Gyr for different values ($\nu$,$\mu$) for a bimodal IMF. The horizontal lines indicate solar and 2.5 times solar metallicities.}
\end{figure}

The ($V-K$)-Mg$_{2}$ diagram is a useful diagnostic, and helps us to check which solutions are reasonable. However, other diagrams such as $U-V$ vs. H$\beta$ 
serve the same purpose (see Figs.~15 and 16 or Figs.~21 and 22). In Figs.~20, 21 and 22 we plot the diagrams for the models with $\nu$ and $\mu_{0}$ which reach a metallicity of 2.5 times solar at $t_{0}$ following Table~8. Focusing our 
attention on the $V-K$ vs. Mg$_{2}$ plots we clearly see that with a bimodal IMF our data can be fitted better. In these plots we also find that better fits can be obtained for $t_{0} \leq 1~Gyr$. The reason for that is clear: if we extend 
the initial period of time we need lower SFR coefficients implying that the most recent stellar populations become relatively more important and therefore the Mg$_{2}$ decreases. The highest values for the Mg$_{2}$ index are in fact 
obtained for $t_{0}=0.2~Gyr$. In Fig.~23 we detail the different contributions to the present time of some of the most representative fits. We see that the contribution of the stars with metallicities lower than solar,
 formed during the very short early stage of the galaxy evolution, is almost negligible (lower than $\sim5\%$). To illustrate this, we have also compared these results with a simple analytic model which yields the same final 
metallicity as our variable IMF model. In the same figure we also plotted an {\em analytical infall model}, in which metal-free gas enters at a rate assumed to equal the stellar birthrate (Larson 1972, Audouze \& Tinsley 1976 for a 
review). We see that this model yields the same results as our numerical variable IMF model. Therefore there are two main ways to obtain the required
metallicity distributions. We have chosen a variable IMF scenario rather than
infall in our numerical models because this would seem to be a more reasonable
physical assumption for ellipticals and bulges, whereas long-term infall would
clearly be a reasonable assumption for the evolution of the disk population such
as that of the solar neighborhood.    
\begin{figure}
\plotone{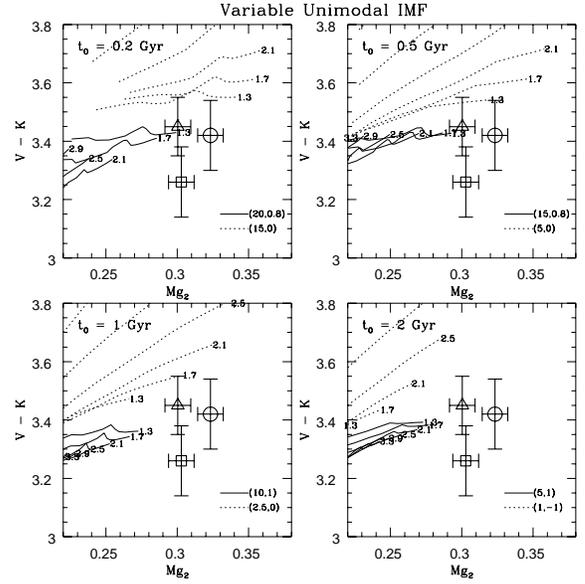}
\caption{The ($V-K$)-Mg$_{2}$ diagram obtained using the assumption that during an initial short period of time (0.2, 0.5, 1 and 2 Gyr.) the IMF was less steep 
(with an slope of $\mu_{0}$) than during the remaining time (with an slope of $\mu$). A unimodal IMF was used. ($\nu$,$\mu_{0}$) are given in brackets while $\mu$ is shown at the end of each curve, which represents the synthetic values 
obtained for assumed ages ranging along the curves from 8 to 17~yr. Notice that the scale is different from that in Figs.~15, 16 and 17. Following Table~8, values of ($\nu$,$\mu_{0}$) were selected to obtain a metallicity around 2.5 times solar 
at the end of the initial period of time. On the diagrams we have represented only the results obtained for specific models ($\nu$,$\mu_{0}$) which bracket 
the observed data. Therefore, acceptable fits will be contained for ($\nu$,$\mu_{0}$) values between these two limiting models.}
\end{figure}
\begin{figure}
\plotone{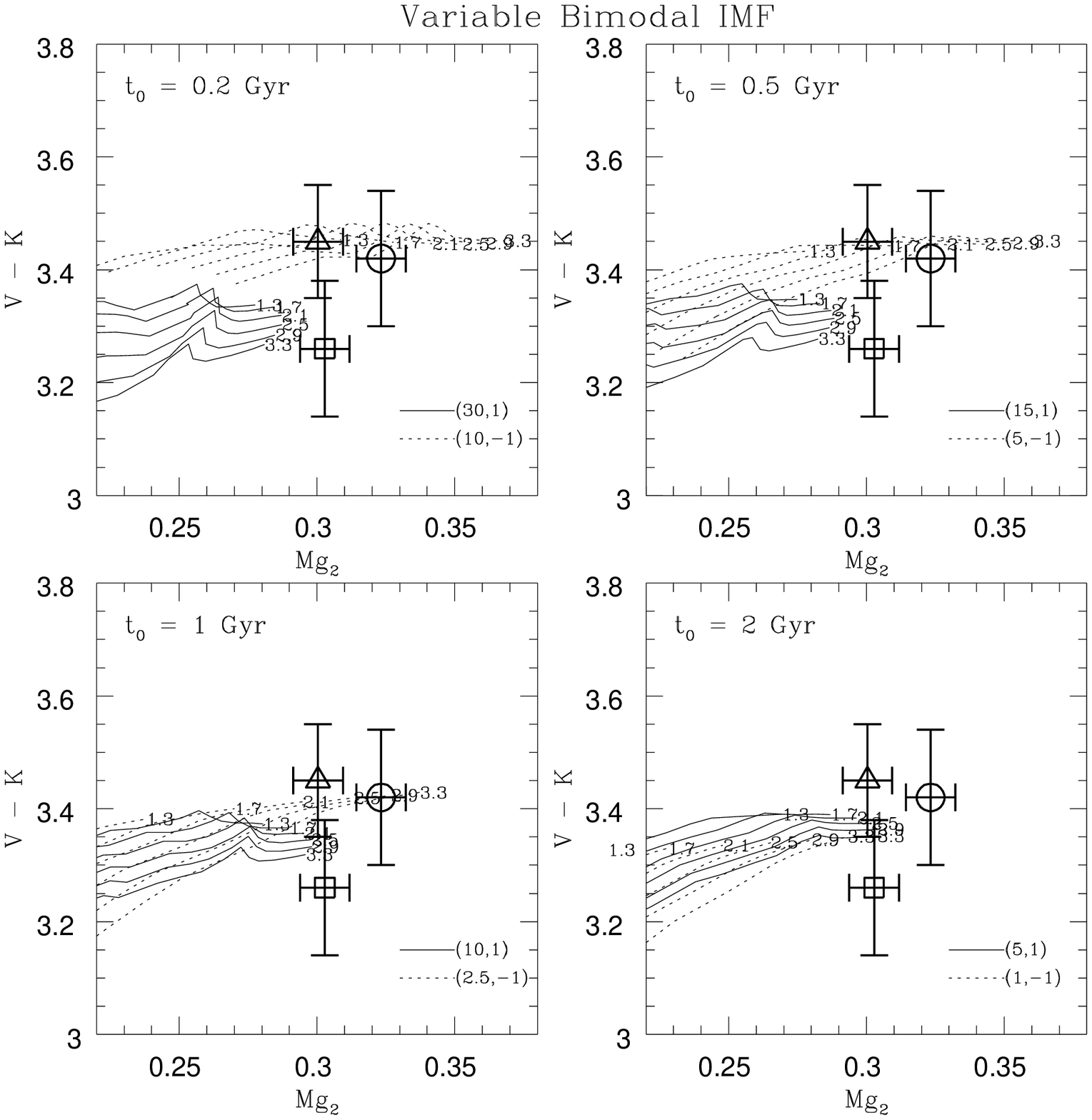}
\caption{The same as in Fig.~20 but for a bimodal IMF. Notice that here the models fit better than in the previous unimodal case.}
\end{figure}
\begin{figure}
\plotone{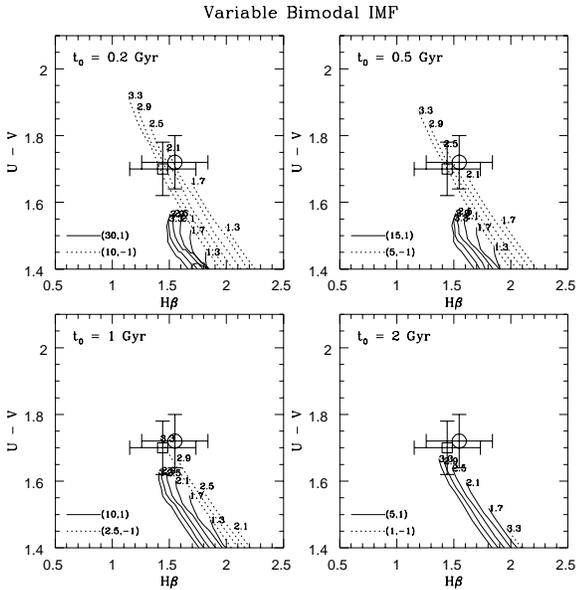}
\caption{The H$\beta$-(U-V) diagram corresponding to the same set of parameters of Fig.~21.}
\end{figure}
\begin{figure}
\plotone{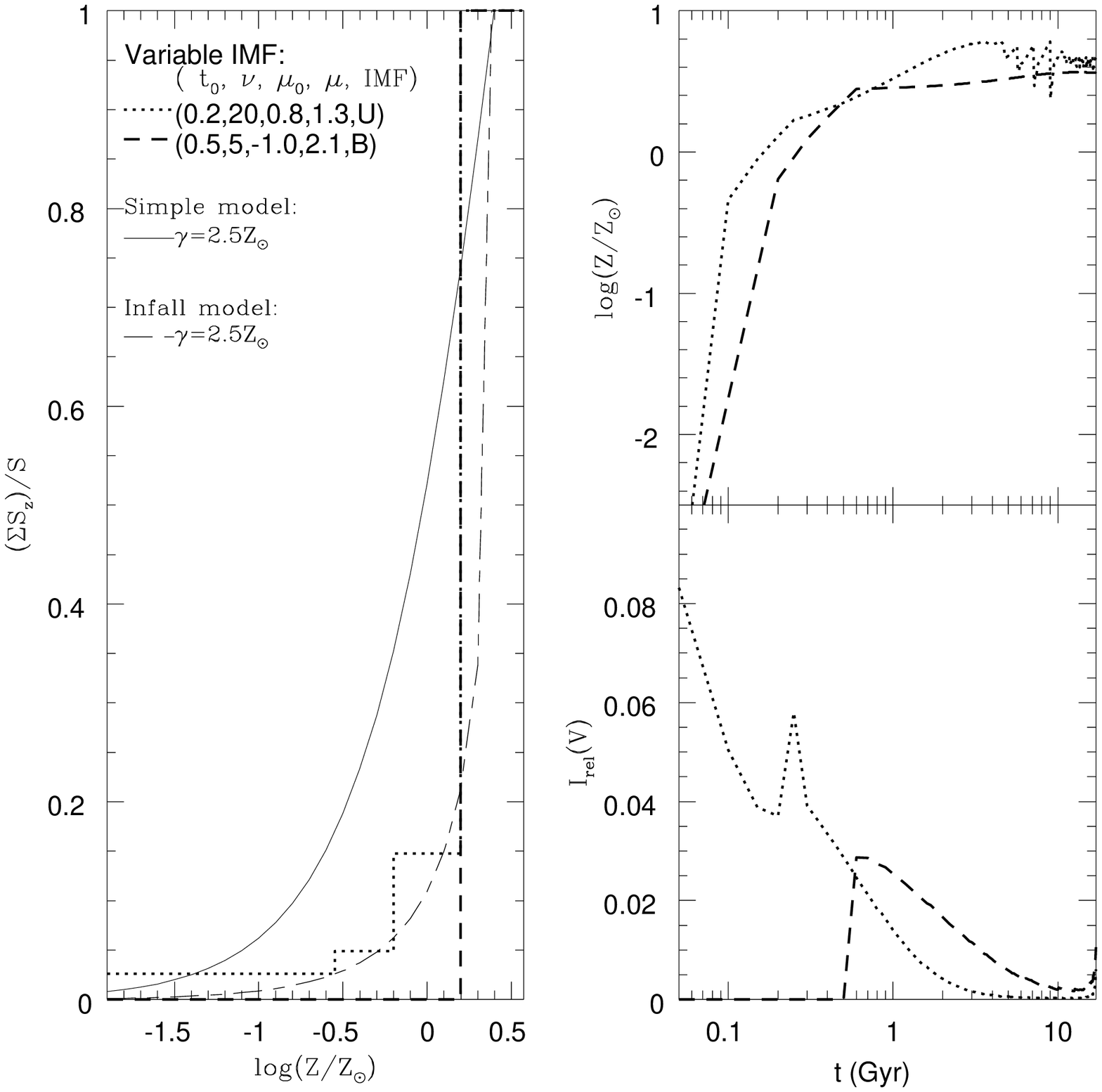}
\caption{The same plots as in Fig.~23 but now using two of the most representative fits of the variable IMF models which fit the data, by inspection of Figs.~20, 21 and 22. For comparison we represent a simple analytical 
model with a final metallicity of $2.5~Z_{\odot}$. Notice that the cumulative fraction of stars with metallicities lower than solar is now below $\sim5\%$. Also notice that the contribution to the present light of the stars born at the early stages of the galaxy evolution (with low metallicities) is much less 
important than in Fig.~18. Finally, we see that another possible scenario capable of explaining the observed distribution of stars is by means of an infall model in which metal-free gas enters at a rate assumed to equal the stellar birthrate (Larson 1972).}
\end{figure}

Finally, since the Mg$_{2}$ should decrease if we choose models ($\nu$,$\mu_{0}$) which reach solar metallicity at $t_{0}$ we also took $t_{0}=0.2~Gyr$ to make Mg$_{2}$ as high as possible. Fig.~24 shows that for 
these models we cannot obtain the observed high values. The Sombrero is the only galaxy which seems to be fitted by this level of metallicity for a remaining $\mu$ lower than 1.7, but this is not a real fit, because with these IMF slopes 
the metallicity continues increasing to reach values higher than Z=0.1 and even more, and therefore the resulting observables are not fitted. NGC~3379 seems to require higher metallicity than solar but not as high as 2.5 times solar, while NGC~4472 seems to be well fitted by a value of 2.5 times solar.

Here we should direct our attention to the recent paper of Gibson (1996) in which he criticizes this variable IMF scenario. He constructs a coupled photometric and chemical evolution package based on the predictions for the 
bimodal star formation/IMF scenario of Elbaz {\it et al.} (1995) implementing both the isochrones of W94 and those of BBCFN for single stellar population models. His models produce $V-K$ colors $\sim0.5 \rightarrow 1.4$ magnitudes too 
red compared to data from ellipticals. We have shown in this section that we can fit the data very well. In our opinion the reason for the discrepancy found by Gibson (1996) is not due to a variable IMF scenario but to the following 
effects: i) our method of converting the isochrone parameters to colors gives a bluer $V-K$, as can be seen for example in Fig.~3 for a single stellar population model of 12~Gyr and metallicity higher than solar. ii) for his calculations 
Gibson (1996) used a SSP while we have followed the chemical evolution during the whole history of the galaxy using our evolutionary model.
\clearpage

\section{Conclusions}
We have developed a stellar population model to apply to early-type galaxies.
It produces optical and near-infrared colors, and absorption line strengths in
lines from 4100 - 8800~\AA~on the Lick system, is applicable to systems of
intermediate or old age, and metallicity larger than 0.1 Z$_\odot$. The model is
chemo-evolutionary, i.e. it calculates the properties of a stellar system,
starting from a primordial gas cloud. However, it can easily be used to predict
the properties of systems with a single age and metallicity. The model colors
and line strengths are determined by integrating stellar observables along
theoretical isochrones, in this way obtaining Single Stellar Populations (SSPs),
and then integrating these SSPs over time. The model uses isochrones with solar
metal abundance ratios. As far as possible empirical calibrations have been 
used to convert theoretical isochrone values to observables for individual
stars. Some properties of the models are:
\begin{figure}
\plotone{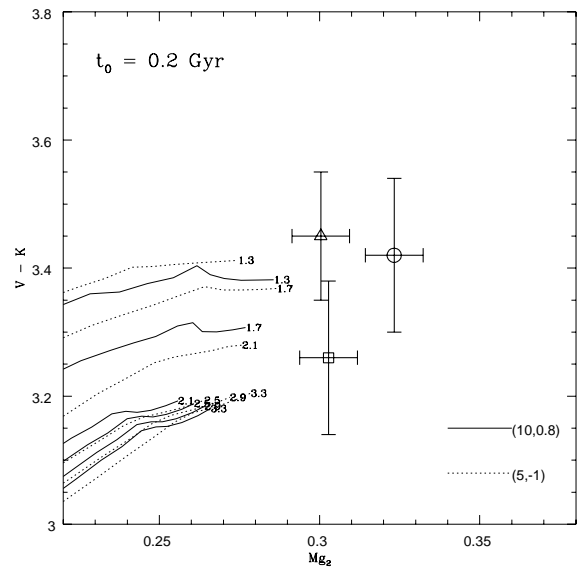}
\caption{The ($V-K$)-Mg$_{2}$ diagram for models which reach solar metallicity
in an initial period of 0.2~Gyr. for a bimodal IMF. We can see that the Mg$_{2}$
index is always lower than the observational data. As can be deduced from 
Fig. ~21 the Mg$_{2}$ index is expected to decrease with increasing $t_{0}$. The metallicities in the three galaxies must therefore clearly be higher than solar.}
\end{figure}

\begin{itemize}
\item Extensive comparisons with models from other authors show
that there is broad overall agreement in the colors and line strengths.
We have also made independent estimates of the errors in
our observables.
\item Our chemical evolutionary model is in good agreement with
Arimoto \& Yoshii (1986). We present optical and IR colors which are likely to be improved as a result of new calibration relations. 
Also, for the first time, we give integrated line strengths for an
evolutionary model.
\item We find that for a closed box approximation the most metal 
rich elliptical galaxies cannot be fitted with a single IMF that is
constant in time. To solve this problem, we propose a scenario invoking
an IMF skewed towards high-mass stars during a short, initial period 
(smaller than 1 Gyr), followed by preferential low-mass star formation in the remaining time.
\item We have briefly tested the model here by fitting it to a few key colors
and line strength indices, using a new data set for 3 standard early-type
galaxies. In general, satisfactory results are reported. A much more comprehensive comparison of theory with observations is given in Paper II.

\end{itemize}

To conclude, we need to make a few statements about the applicability of this 
model. Given the nature of this type studies, the numbers given in the text
will soon cease to be the most accurate possible, because better
isochrones become available, or better calibrations linking one parameter
to another. Therefore, we will try to update the model as time goes on,
and interested people can always obtain the most recent version
electronically from the authors. There are however a few areas in
which we think that further effort by the astronomical community is needed
to improve models of this kind. These are:

\begin{itemize}
\item We show that absorption line strengths are generally more 
accurate than integrated colors. To change this situation better 
color-color relations are needed, especially for very high and
very low metallicities.
\item Inclusion of more absorption line observations will make this kind of
models more useful. Especially in the near-UV or near-IR very little work
has been done, except in the region of the Ca II triplet. More libraries
like the one of Worthey {\it et al.} (1994) are urgently needed.
\item It looks as if a substantial part of the uncertainties are due to 
incorrect treatment of the later stages of stellar evolution, such as the AGB.
Theoretical work in this area would be very valuable.
\end{itemize}

\acknowledgments
We are grateful to Franco Fagotto for discussions and for his valuable comments and suggestions on the manuscript, which have greatly improved the final version. We also would like to thank the referee, Guy Worthey, for improving substantially the quality of this paper. We are grateful to the Padova group for providing us with the set of theoretical isochrones. We are indebted to Onno Pols who kindly gave us the calculations for the very low-mass stars, and to Angel Alonso, 
Santiago Arribas and Carlos Mart\'{\i}nez who have made available their observational results before publication. We also thank Gustavo Bruzual and St\'ephane Charlot who provided us with their latest calculations of integrated colors and absorption line indices. We also thank Nobuo Arimoto, Marijn Franx and Bianca Poggianti for useful comments on the manuscript. This work was partially supported by grants
PB91-0510 and PB94-1107 of the Spanish DGICYT. A.V. thanks the Kapteyn Institute in Groningen for financial assistance during a working visit.


\begin{references}
\reference Alexander, D.R. \& Ferguson, J.W., 1994, \apj, 473, 879
\reference Alongi, M., Bertelli, G., Bressan, A., Chiosi, C., Fagotto, F., Greggio, L., Nasi, E., 1993, \aaps, 97, 851
\reference Alonso, A., Arribas, S. \& Mart{\'{\i}}nez-Roger, C., 1994, \aaps, 107, 365
\reference Alonso, A., Arribas, S. \& Mart{\'{\i}}nez-Roger, C., 1995, \aap, 297, 197
\reference Alonso, A., Arribas, S. \& Mart{\'{\i}}nez-Roger, C., 1996, \aap, In press
\reference Arimoto, N., \& Yoshii, Y., 1986, \aap, 164, 260 ({\bf AY86})
\reference Arimoto, N., \& Yoshii, Y., 1987, \aap, 173, 23
\reference Arnaud, M., Rothenflug, R., Boulade, O., Vigroux, L. \& Vangioni-Flam, E., \aap, 254, 49
\reference Audouze, J. \& Tinsley, B.M., 1976, \aapr, 14, 43
\reference Bertelli, G., Bressan, A., Chiosi, C., Fagotto, F. \& Nasi, E., 1994, \aaps, 106, 275 ({\bf BBCFN})
\reference Bessell, M.S., 1983, \pasp, 95, 480
\reference Bessell, M.S., 1990, \pasp, 102, 1181
\reference Bessell, M.S., Wood, P. R., 1984, \pasp, 96, 247
\reference Bessell, M.S. \&  Brett, J.M., 1988, \pasp, 100, 1134
\reference Bessell, M.S., Brett, J.M., Scholz, M. \& Wood, P.R., 1989, \aaps, 77, 1
\reference Bessell, M.S., Brett, J.M., Scholz, M. \& Wood, P.R., 1991, \aaps, 89, 335
\reference Bica, E., 1988, \aap, 195, 76
\reference Bica, E., Pastoriza, M., Maia, M., da Silva, L. \& Dottori, H., 1991, \aj, 102, 1702
\reference Bica, E., Clar\'\i a, J.J \& Dottori, H., 1992, \aj, 103, 1859
\reference Bressan, A., Chiosi, C., Fagotto, F., 1994, \apjs, 94, 63
\reference Bressan, A., Chiosi, C. \& Tantalo, R., 1996, \aap, submitted
\reference Bruzual, G., 1983, \apj, 273, 205
\reference Bruzual, G., 1992, in The Stellar Populations of Galaxies, ed. B. Barbuy \& A. Renzini (Dordrecht:Kluwer), 311
\reference Bruzual, G. \& Charlot, S., 1993, \apj, 405, 538
\reference Bruzual, G. \& Charlot, S., 1996, In preparation
\reference Burstein, D., Faber, S. M., Gaskell, C. M. \& Krumm, N., 1984, \apj, 287, 586
\reference Buser, R. \& Kurucz, R., 1978, \aap 70, 555
\reference Buzzoni, A., 1989, \apjs, 71, 817
\reference Caldwell, N., Kennicutt, R. \& Schommer, R., 1994, \aj, 108, 1186
\reference Casuso, E., 1991, Ph. D. Thesis, Univ. of La Laguna
\reference Casuso, E., Vazdekis, A., Peletier, R. \& Beckman, J.E., 1996, \apj, 458, 533 
\reference Charlot, S. \& Bruzual, G., 1991, \apj, 367, 126
\reference Charlot, S., Worthey, G. \& Bressan, A., 1996, \apj, 457, 625
\reference Clayton, D.D., 1985, in Nucleosynthesis, Univ. Chicago Press, W. D. Arnett \& Truran, eds., p. 65
\reference Clayton, D.D., 1986, \pasp, 98, 968
\reference Code, A.D., Davis, J., Bless, R.C., \& Hanbury Brown, R., 1976, \apj, 203,417
\reference Covino, S., Galletti, S. \& Pasinetti, L.E., 1995, \aap, 303, 79
\reference D{\'\i}az, A., Terlevich, E. \& Terlevich, R., 1989, \mnras, 239, 325
\reference Eggleton, P.P., 1973, \mnras, 163, 279
\reference Elbaz, D., Arnaud, M \& Vangioni-Flam, E., 1995, \aap, 303, 345
\reference Faber, S.M., 1972, \aap, 20, 361.
\reference Faber, S.M., Friel, E.D., Burstein, D., Gaskell, C.M., 1985, \apjs, 57, 711
\reference Fluks, M.A., Plez, B., Th\'e, P.S., de Winter, D., Westerlund, B.E. \& Steenman, H.C., 1994, \aaps, 105, 311
\reference Frogel, J.A., Persson, S.E. \& Cohen, J.G., 1981, \apj 246, 842
\reference Gibson, B.K., 1996, \mnras, submitted
\reference Girardi, L., Chiosi, C., Bertelli, G. \& Bressan, A., 1995, \aap, 298, 87
\reference Gorgas, J., Faber, S.M., Burstein, D., Gonzalez, J.J., Courteau, S. \& Prosser, C., 1993, \apjs, 86, 153
\reference Guiderdoni, B., Rocca-Volmerange, B., 1987, \aap, 186, 1
\reference Guiderdoni, B., Rocca-Volmerange, B., 1990, \aap, 227, 362
\reference Guiderdoni, B., Rocca-Volmerange, B., 1991, \aap, 252, 435
\reference Han, Z., Podsiadlowski, P. \& Eggleton, P.P., 1994, \mnras, 270, 121
\reference Hes, R. \& Peletier, R., \aap, 268, 539
\reference Huebner, W.F., Merts, A.L., Magee, N.H. \& Argo, M.F., 1977, Los Alamos Sci. Lab. Rep. LA-6760-M
\reference Houdashelt, M.L., Frogel, J.A. \& Cohen, J.G., 1992, \aj 103, 163
\reference Iglesias, C.A., Rogers, F.J., Wilson, B.G., 1992, \apj, 397, 717
\reference Johnson, H.L., 1966, \araa, 4, 193
\reference Kennicutt, R.C., 1989, \apj, 344, 685
\reference Kroupa, P., Tout, C.A. \& Gilmore, G., 1993, \mnras, 262, 545
\reference Kurucz, R.L., 1992, in The Stellar Populations of Galaxies,ed. B. Barbuy \& A. Renzini (Dordrecht: Kluwer), 225
\reference Lacey, G., Fall, S.M., 1985, \apj, 290, 154
\reference Lacey, C.G., Guiderdoni, B., Rocca-Volmerange, B. \& Silk, J., 1993, \apj, 402, 15
\reference Larson, R.B., 1972, Nature Phys. Sci., 236, 7
\reference Larson, R.B. \& Tinsley, B.M., 1978, \apj, 219, 46
\reference Maeder,A., 1992, \aap, 264, 105
\reference Matteucci, F., Tornambe, A., 1987, \aap, 185, 51
\reference O'Connell, R.W., 1976, \apj, 206, 370
\reference O'Connell, R.W., 1980, \apj, 236, 430
\reference Pagel, B.E.J., 1989, in Evolutionary Phenomena in Galaxies, ed. J. Beckman \& B.E.J. Pagel (Cambridge: Cambridge Univ. Press), 201
\reference Peletier, R.F., 1989, Ph. D. Thesis, Univ. of Groningen
\reference Peletier, R.F., Davies, R.L., Illingworth, G.D., Davis, L.E. \& Cawson, M., 1990a, \aj, 100, 1091
\reference Peletier, R.F., Valentijn, E.A. \& Jameson, R.F., 1990b, \aap, 233, 62
\reference Pickles, A.J., 1985, \apj, 296, 340
\reference Pols, O.R., Tout, C.A., Eggleton, P.P. \& Han, Z., 1995, \mnras, accepted
\reference Reimers, D., 1975, Mem. Soc. R. Sci. Liege, Ser. 6, Vol. 8, 369.
\reference Renzini, A. \& Voli, M., 1981, \aap, 94, 175
\reference Renzini, A. \& Buzzoni, A., 1986, in Spectral Evolution of Galaxies, ed. C. Chiosi \& A. Renzini (Dordrecht: Reidel), 195
\reference Ridgway, S.T., Joyce, R.R., White, N.M. \& Wing, R.F., 1980, \apj, 235, 126
\reference Rieke, G.H. \& Lebofsky M.J., 1985, \apj, 288, 618
\reference Rocca-Volmerange, B., 1989, \mnras, 236, 7
\reference Rocca-Volmerange, B. \& Guiderdoni, B., 1987, \aap, 175, 15
\reference Rocca-Volmerange, B. \& Guiderdoni, B., 1988, \aaps, 75, 93
\reference Rocca-Volmerange, B. \& Guiderdoni, B., 1990, \mnras, 247, 166
\reference Scalo, J.M., 1986, Fund. Cosmic Phys., 11, 1.
\reference Schmidt, M., 1959, \apj, 129, 243 
\reference Schmidt, M., 1963, \apj, 137, 758 
\reference Searle, L., Sargent, W.L.W. \& Bagnuolo, W.G., 1973, \apj, 179, 427
\reference Searle, L., Wilkinson, A. \& Bagnuolo, W.G., 1980, \apj, 239, 803
\reference Spinrad, H. \& Taylor, B.J., 1971, \apjs, 22, 445
\reference Stetson,P.B. \& Harris, W.E., 1988, \aj, 96, 909
\reference Tantalo, R., Chiosi, C. Bressan, A. \& Fagotto, F., 1995, \aap, submitted
\reference Terndrup, D.M., Frogel, J.A. \& Withford,A.E., 1990, \apj, 357, 453
\reference Terndrup, D.M., Frogel, J.A. \& Withford, A.E., 1991, \apj 378, 742
\reference Tinsley, B.M., 1968, \apj, 151, 547
\reference Tinsley, B.M., 1972,\aap, 20, 383
\reference Tinsley, B.M. \& Gunn, J.E.,1976, \apj, 203, 52
\reference Tinsley, B.M., 1978a, \apj, 220, 816
\reference Tinsley, B.M., 1978b, \apj, 222, 14
\reference Tinsley, B.M., 1980, Fund. Cosmic Phys., 5, 287
\reference Theis, C., Burkert, A., Hensler, G., 1992, \aap, 265, 465
\reference Tosi, M. 1988, \aap, 197, 33
\reference Turnrose, B.E., 1976, \apj, 210, 33
\reference Twarog, B.A., 1980, \apj, 242, 242
\reference VandenBerg, D.A., 1983, \apjs, 51,29
\reference VandenBerg, D.A., 1992, \apj, 391, 685
\reference Vazdekis, A., Peletier, R. F., Beckman, J. E. \& Casuso, E., 1996, \apj, submitted ({\bf Paper II})
\reference Weiss, A., Peletier, R.F. \& Matteucci, F., 1995, \aap, 296, 73
\reference Whitford, A.E., 1978, \apj, 226, 777
\reference Worthey, G., Faber, S.M. \& Gonzalez, J.J., 1992, \apj, 398, 69
\reference Worthey, G., Faber, S., Gonz\'alez, J. \& Burstein, D., 1994, \apjs, 94, 687
\reference Worthey, G., 1994, \apjs, 95, 107 ({\bf W94})
\reference Wu, C.C., Faber, S.M., Gallagher, J.S., Peck, M. \& Tinsley, B.M., 1980, \apj, 237, 290
\reference Yoshii, Y. \& Takahara, F., 1988, \apj, 299, 593
\end{references}
\end{document}